\documentclass[twocolumn]{aastex63}
\usepackage{diagbox}
\usepackage{multirow}
\usepackage{amsmath}
\usepackage{lineno}

\begin{document}

\title{Constraints on the cosmological parameters with three-parameter correlation of Gamma-ray bursts}
\author{Jia-Lun Li}
\author{Yu-Peng Yang}
\author[0000-0003-0672-5646]{Shuang-Xi Yi$^{\dag}$}
\affiliation{School of Physics and Physical Engineering, Qufu Normal University, Qufu 273165, China; yisx2015@qfnu.edu.cn,ypyang@qfnu.edu.cn}
\author{Jian-Ping Hu}
\author{Fa-Yin Wang}
\affiliation{School of Astronomy and Space Science, Nanjing University, Nanjing 210023, China; fayinwang@nju.edu.cn}
\author{Yan-Kun Qu}
\affiliation{School of Physics and Physical Engineering, Qufu Normal University, Qufu 273165, China; yisx2015@qfnu.edu.cn,ypyang@qfnu.edu.cn}

\begin{abstract}

As one of the most energetic and brightest events, gamma-ray bursts (GRBs) can be treated as a promising probe of the high-redshift universe.
Similar to type Ia supernovae (SNe Ia), GRBs with same physical origin could be treated as standard candles.
We select GRB samples with the same physical origin, which are divided into two groups.
One group is consisted of 31 GRBs with a plateau phase feature of a constant luminosity followed by a decay index of about -2
in the X-ray afterglow light curves, and the other has 50 GRBs with a shallow decay phase in the optical light curves.
For the selected GRB samples, we confirm that there is a tight correlation between the plateau luminosity $L_0$,
the end time of plateau $t_b$ and the isotropic energy release $E_{\gamma,iso}$.
We also find that the $L_0-t_b-E_{\gamma,iso}$ correlation
is insensitive to the cosmological parameters and no valid limitations on the cosmological parameters
can be obtained using this correlation.
We explore a new three-parameter correlation $L_0$, $t_b$, and the spectral peak energy in the rest frame $E_{p,i}$ ($L_0-t_b-E_{p,i}$),
and find that this correlation can be used as a standard candle to constrain the cosmological parameters.
By employing the optical sample only, we find the constraints of $\Omega_m = 0.697_{-0.278}^{+0.402}(1\sigma)$ for
a flat $\Lambda$CDM model.
For the non-flat $\Lambda$CDM model, the best-fitting results are $\Omega_m = 0.713_{-0.278}^{+0.346}$, $\Omega_{\Lambda} = 0.981_{-0.580}^{+0.379}(1\sigma)$. For the combination of the X-ray and optical smaples, we find $\Omega_m = 0.313_{-0.125}^{+0.179}(1\sigma)$
for a flat $\Lambda$CDM model, and $\Omega_m = 0.344_{-0.112}^{+0.176}$, $\Omega_{\Lambda} = 0.770_{-0.416}^{+0.366}(1\sigma)$
for a non-flat $\Lambda$CDM model.
\end{abstract}


\section{introduction}

Gamma-ray bursts (GRBs) are among the most energetic explosive events with a luminosity $L \sim 10^{47}$ - $10^{54}$
$\rm erg~s^{-1}$ in the Universe (\citealp{1973ApJ...182L..85K}; \citealp{2006RPPh...69.2259M}; \citealp{2009ARA&A..47..567G}; \citealp{2011A&A...536A..96W};  \citealp{2015PhR...561....1K}). Based on the distribution of the bimodal duration time $T_{90}$, GRBs are typically divided into two categories of short GRBs (SGRBs, $T_{90}<2s$) and long GRBs (LGRBs, $T_{90}> 2s$)~(\citealp{1993ApJ...413L.101K, 2013ApJ...763...15Q}).
The LGRBs most likely originate from the collapses of the massive stars~(\citealp{1993ApJ...405..273W, 1998ApJ...494L..45P, 1999ApJ...524..262M, 2001ApJ...550..410M}).
The progenitors of SGRBs stem from the merger of two neutron stars (NS) or a NS and a black hole (BH) in the system of binary~(\citealp{2006ARA&A..44..507W, 2009ARA&A..47..567G, 2017ApJ...848L..13A,2021ApJ...922..255T}).
At present, the maximum redshift of GRB observed is $z \sim 9.4$~(\citealp{2011ApJ...736....7C}), and
they could be detected up to $z \sim 20$~(\citealp{2000ApJ...536....1L}). Therefore, GRBs can be used to
investigate the characters of the Universe at high-redshift~(\citealp{2004ApJ...612L.101D,  2004ApJ...613L..13G, 2006MNRAS.369L..37L, 2007ApJ...660...16S, 2008MNRAS.391L...1K, 2011MNRAS.415.3423W, 2013IJMPD..2230028A,2017IJMPD..2630002W, 2018SSPMA..48c9505W, 2021JCAP...09..042K,2022MNRAS.514.1828D,2022PASJ...74.1095D}).
On the other hand, SNe Ia~(\citealp{1993ApJ...413L.105P, 1998AJ....116.1009R, 1999ApJ...517..565P}) and the cosmic microwave background (CMB)~(\citealp{2003ApJS..148..175S, 2014A&A...571A..16P, 2016A&A...594A..13P, 2020A&A...641A...6P})
have been successfully used as cosmological probes.
Due to the physical mechanism of SNe Ia, its maximum luminosity is limited, leading to the detected upper limit
of redshift is not very large. The CMB provides relevant information about the early universe.
From this point of view, GRBs are regarded as a complement to SNe Ia and CMB. Moreover, gamma-ray photons are largely unaffected by the interstellar medium (ISM) that SNe Ia faces as they travel towards us~(\citealp{2015NewAR..67....1W}).

Similar to SNe Ia, GRBs can be standardized to be a cosmic distance indicator for cosmological purposes
by using correlations between their observable quantities (\citealp{2022MNRAS.516.1386C, 2022MNRAS.512..439C, 2022MNRAS.510.2928C, 2022MNRAS.516.2575J, 2022ApJ...941...84L, 2022ApJ...935....7L,2023MNRAS.tmp..889L}).
In general, the relations can be divided into three main categories: (1) the correlations observed in the prompt phase of the
emission including
the Amati correlation~(\citealp{2002A&A...390...81A}), the Ghirlanda correlation~(\citealp{2004ApJ...616..331G}), and the Yonetoku correlation~(\citealp{2004ApJ...609..935Y});
(2) the afterglow correlations involving the Dainotti correlation of $L_0-t_b$~(\citealp{2008MNRAS.391L..79D}),
(3) the prompt-afterglow correlations including Liang-Zhang correlation~(\citealp{2005ApJ...633..611L}).
There have been extensive investigations on the Dainotti relation, which is a correlation between the plateau luminosity $L_0$ and the end time of the plateau $t_b$. Moreover, there are many studies that have used the Dainotti relation to measure cosmological parameters~(\citealp{2009MNRAS.400..775C, 2010MNRAS.408.1181C, 2013MNRAS.436...82D, 2014ApJ...783..126P, 2015A&A...582A.115I, 2022ApJ...925...15L}). \cite{2016A&A...585A..68W} have used
the Dainotti relation to standardize the afterglow light curves of long GRBs,
and the GRB samples are divided into gold and silver sample groups according to the behaviors of the light curves.
\cite{2021MNRAS.507..730H} conducted updated investigations with
SGRBs whose spin-down is dominated by magnetic dipole (MD) radiations (MD-SGRBs) and LGRBs whose spin-down is dominated by gravitational wave (GW) emission (GW-LGRBs). The method has been proposed by~\cite{2012A&A...538A.134X} to extend the Dainotti relation by adding the isotropic energy $E_{\gamma,iso}$ into the $L_0-t_b$ relation and found a much tighter correlation than that of two-parameter.
Here we will use the relation $L_0-t_b-E_{\gamma,iso}$ to probe the cosmological parameters.
Moreover, since the correlation between $E_{\gamma,iso}$ and the spectral peak energy in the rest frame $E_{p,i}$~(\citealp{2002A&A...390...81A}), we will also
investigate whether $L_0-t_b-E_{p,i}$ correlation can be used to probe the cosmological parameters.
\cite{2021ApJ...920..135X} have used the relations $L_0-t_b-E_{\gamma,iso}$ and $L_0-t_b-E_{p,i}$ to constrain the cosmological parameters, based on a sample including 121 long GRBs. We will study these two correlations by employing different samples.

If the central engine of a GRB is powered by a newly born fast spinning neutron star with high magnetic field,
the energy injection from the magnetar could cause the plateau phase in the X-ray light curves~(\citealp{1998A&A...333L..87D, 2001ApJ...552L..35Z, 2011MNRAS.413.2031M}). Based on the observations of Swift~(\citealp{2004ApJ...611.1005G}),
a significant fraction of GRBs shows a plateau phase in the X-ray light curves
followed by a decay phase in afterglows~(\citealp{2006ApJ...642..354Z, 2006ApJ...642..389N, 2006ApJ...647.1213O, 2007ApJ...670..565L, 2015ApJ...807...92Y,2016ApJS..224...20Y,2023ApJS..265...56L}).
This characteristic of the light curve could be explained by the possibility that the energy injection from the magnetar causes a shallow decline phase in afterglows (\citealp{1998A&A...333L..87D, 2001ApJ...552L..35Z}). \cite{1998A&A...333L..87D} have suggested that the rotational energy of a newly born magnetar is released as the gravitational wave and electromagnetic radiation,
leading to the neutron star spin down. If the spin down is dominated by magnetic dipole radiation (MD-radiation),
the corresponding luminosity evolving with time can be written as~(\citealp{1998A&A...333L..87D})

\begin{equation}\label{eq2}
L=L_0\times \frac{1}{(1+t/t_b)^2}\simeq\left\{
\begin{array}{lcl}
L_{0} & & {t\ll t_b} ,\\
L_{0}{(t/t_b)}^{-2} & & {t\gg t_b},
\end{array}
\right.
\end{equation}
where $L_0$ and $t_b$ represent the characteristic spin-down luminosity and end time of the plateau, respectively.
The values of these parameters can be obtained by fitting the X-ray afterglow light curves with the plateau phase.

In recent years,
SNe Ia have been studied extensively as a well-established class of standard candle, since SNe Ia with the same source of systematics have a nearly uniform luminosity with an absolute magnitude $M\simeq-19.5$ (\citealp{2001LRR.....4....1C}).
Similar to SNe Ia, GRBs with plateau phase caused by the same physical mechanism should be standardized as standard candles.

Recently, GRBs have been classified in more detail with different decay indices~(\citealp{2022ApJ...924...97W}).
These classified GRBs are then standardized using Dainotti correlation, and a more tight cosmological constraints are obtained.
According to the observations of Swift, many long-GRBs have a trait of plateau phase and a normal decay phase.
Additionally, an analogous shallow decay phase also appears in the optical afterglow of GRBs.
These shallow decay phase of optical light curves should have a similar physical mechanism.
Therefore, similar to the case of X-ray, the classification of optical light curves is also based on the decay indices.

\cite{2018ApJ...863...50S} have found that
the correlations of $L_0-t_b-E_{\gamma,iso}$ or $L_0-t_b-E_{p,i}$ are tighter
than that of $L_0-t_b$. In this work, we will investigate the selected X-ray and optical samples,
whose spin-down is potentially dominated by the same physical mechanism, and standardize them using
the correlation of $L_0-t_b-E_{\gamma,iso}$ and $L_0-t_b-E_{p,i}$. In the next section, we will briefly introduce the sample selection.
In Sec. III, we standardize X-ray and optical samples utilizing the correlations of $L_0-t_b-E_{\gamma,iso}$ and $L_0-t_b-E_{p,i}$.
In Sec. IV, we use the standardized samples to constrain the cosmological parameters for $\Lambda$CDM model,
and discuss which correlation is better for probing the Universe at higher redshifts.
The conclusions are given in Sec. V.

\section{Sample selection of GRBs}

As previously stated, if the central engine of a GRB is powered by a newly born magnetar, the continuous energy injection from magnetar will cause a plateau phase in the X-ray light curves~(\citealp{1998A&A...333L..87D, 2001ApJ...552L..35Z, 2011MNRAS.413.2031M}). The energy reservoir of newly born magnetized neutron star is rotational energy, and the spin-down of a newly born magnetar is through a combination of the electromagnetic dipole radiation and gravitational wave emission. If the spin-down of newly born magnetar is dominated by MD-radiation, the light curves of X-ray afterglow will show a plateau followed by a normal decay phase
with a decay index of about -2. On the other hand, if the spin-down is dominated by GW-emission, the decay index
is about -1 for the X-ray afterglow light curves.

These predictions have been confirmed by the Swift observations, and
part of GRBs are characterized by a plateau phase followed by a normal decay phase in the early X-ray afterglow.
\cite{2019ApJS..245....1T} conducted a statistical investigation of 174 GRBs with a plateau phase in the X-ray afterglow.
It has been discovered that a tight correlation between $L_0$ and $t_b$ (Dainotti relation)
(\citealp{2008MNRAS.391L..79D}) can be used to measure the cosmological parameters by investigating the GRBs with X-ray afterglow plateau phases~(\citealp{2009MNRAS.400..775C, 2010MNRAS.408.1181C, 2013MNRAS.436...82D, 2014ApJ...783..126P, 2015A&A...582A.115I}).
Whereas, the results of the cosmological constraints are loose, and the main reason is the samples are not well selected (\citealp{2022ApJ...924...97W}).
Similar to the supernova cosmology, where only type Ia supernova caused by the same physical mechanism can serve as standard candles,
GRBs with an X-ray plateau potentially generated by the same physical mechanism are selected.
The decaying indices of the X-ray light curves could be explained by the loss of rotational energy of magneter in different ways.
Therefore, it is necessary to conduct a more detailed classification of GRBs according to different decay indices.

\cite{2022ApJ...924...97W} and \cite{2021MNRAS.507..730H} have carefully selected
and classified the GRB samples into several sample spaces, MD-SGRBs, MD-LGRBs, and GW-LGRBs according to the decay indices in the X-ray afterglow. Theses X-ray light curves are then standardized using Dainotti relation.
Here we adopt 31 LGRBs selected by~\cite{2022ApJ...924...97W}, which are divided into gold and silver samples.
The X-ray sample is selected using the following criteria.

$\bullet$ There are no weak flares, especially during the plateau.

$\bullet$ There are enough data points at plateau and decay phase, and the data points have good coverage of the light curve.

$\bullet$ There is an obvious plateau in gold sample, and an expected platform phase in silver sample by analyzing the XRT and BAT data.

$\bullet$ The duration time of decay phase is larger than 5$t_b$.

$\bullet$ The plateau phase is followed by a decay index of about -2.

The selected samples can improve the reliability of the fitting,
and are consistent with the fact that the energy injection of magnetar electromagnetic dipole emission is greater than
the external shock emission (the decay index is about -1.2). Moreover, the X-ray afterglow light curves of all samples show a plateau stage with
a constant luminosity followed by a decay stage with a decay index of about -2.

Not only do many GRBs exhibit a plateau phase in X-ray afterglow followed by a normal decay phase,
but a similar decay phase also appear in the optical light curves. However, only a small part of the optical afterglow possesses
a plateau phase~(\citealp{2012ApJ...758...27L}). This particular phase of shallow decay in the optical light curves may originate from the same physical process~(\citealp{1998PhRvL..81.4301D,1998A&A...333L..87D,2001ApJ...552L..35Z, 2006MNRAS.372L..19F, 2007ApJ...670..565L, 2013MNRAS.430.1061R, 2014ApJ...785...74L,2022ApJ...924...69Y}).  \cite{2018ApJ...863...50S} screened 50 GRBs samples from the published literature for the studies of
the correlation of optical plateaus. The selection criteria for optical sample are similar to those for X-rays. The optical sample is selected in terms of the following criteria.

$\bullet$ There is an obvious plateau stage in the optical afterglow curve, where a shallow decay or a slight rising phase is allowed.

$\bullet$ The plateau phase is followed by a normal decay or an even steeper decay (such as GRB 030429).

The selected well-sampled afterglows with obvious plateau stage are the transition of the optical afterglow light curves from a shallow decay (or a slight rising phase) to normal decay (or an even steeper decay). It is worth noting that GRBs 050319, 060526 and 080310 are included in both groups of samples.
We will use these selected 50 optical sample and 31 X-ray sample for our studies.

\section{The three-parameter correlation of X-ray and optical samples}
\subsection{Fitting the correlation}

With the plateau flux $F_0$ obtained, the luminosity $L_0$ of the plateau phase can be written as

\begin{equation}\label{eq:L0}
L_0 = \frac{4{\pi}d_L^2F_0}{\left(1+z\right)},
\end{equation}
where $d_L$ is the luminosity distance. For the X-ray luminosity, K-correction of $(1+z)^{1-\beta}$ should be included,
where $\beta$ is the spectral index of the plateau phase. In the flat universe model, the luminosity distance $d_L$ can be written as

\begin{equation}\label{eq:dl_flat}
d_L = \frac{c\left(1+z\right)}{H_0}{\int_0^z}\frac{dz}{\sqrt{{\Omega_m}{\left(1+z\right)}^3+{\Omega_{\Lambda}}}},
\end{equation}
where ${\Omega_m}$ and ${\Omega_{\Lambda}}$ represent the density parameters of matter and dark energy, respectively.

\cite{2022ApJ...924...97W} have pointed out that for a group of GRBs (e.g., the GRBs with redshift $z<0.1$),
if their distances can be obtained directly by observation, the deriving $L_0 - t_b$ correlation is
model-independent.
The relationship between two parameters $L_0$ and $t_b$ can be written as follows

\begin{equation}\label{eq6}
{\rm{log}}\left(\frac{L_0}{10^{47}\rm{erg~s^{-1}}}\right)=k \times {\rm log}\frac{t_b}{10^3\left(1+z\right)s} + b.
\end{equation}

\cite{2022ApJ...924...97W}  has fitted the $L_0 - t_b$ relation using X-ray sample.
Furthermore, for a group of X-ray plateau samples, a much tighter three-parameter correlation
of $L_0 - t_b - E_{\gamma,iso}$ is obtained with the isotropic energy release $E_{\gamma,iso}$~(\citealp{2012A&A...538A.134X, 2023ApJ...943..126D}).
\cite{2018ApJ...863...50S} studied the relationship of $L_0 - t_b - E_{\gamma,iso}$ by
using the well-sampled optical light curves
of 50 GRBs. Similar to the $L_0 - t_b$ relation, the  $L_0 - t_b - E_{\gamma,iso}$ relation can be expressed as

\begin{equation}\label{eq7}
{\rm log}\frac{L_0}{10^{47}{\rm erg~s^{-1}}} = a + b{\rm log}\frac{t_b}{10^3s} + c{\rm log}\frac{E_{\gamma,iso}}{10^{53}\rm erg},
\end{equation}
where a, b and c can be determined by fitting observed data\footnote{In fact,
a is a constant. b and c are actually the power-law indices of $t_b$ and $E_{\gamma,iso}$ when
we consider $L_0$ as a power law functions of $t_b$ and $E_{\gamma,iso}$}. The isotropic energy of the prompt emission is

 \begin{equation}\label{Eiso}
E_{\gamma,iso}^{'} = \frac{4{\pi}d_L^2S}{1+z},
\end{equation}
where $S$ is the fluence. Due to cosmological time dilaton, the energy bands of the observer frame and rest frame are different.
Therefore, K-correction should be considered in the calculation of $E_{\gamma,iso}$.
Including K-correction $E_{\gamma,iso}$ can be written as

\begin{equation}\label{}
E_{\gamma,iso} = E_{\gamma,iso}^{'}\times\frac{\int_{E_1/1+z}^{E_2/1+z}E\phi(E)dE}{\int_{E_1}^{E_2}E\phi(E)dE},
\end{equation}
where $E_2$ and $E_1$ are the upper and lower limits of the detector energy range, respectively.
$\phi(E)$ is the energy spectrum, which can be modeled with a smoothly broken power law~(\citealp{1993ApJ...413..281B}),

\begin{equation}\label{eq:phiE}
\phi(E)=\left\{
\begin{array}{lcl}
AE^{\alpha}e^{-(2+\alpha)E/E_{p,obs}} &, & {E \leq \frac{\alpha-\beta}{2+\alpha}E_{p,obs}}, \\
BE^{\beta} &, &{E > \frac{\alpha-\beta}{2+\alpha}E_{p,obs}},
\end{array}
\right.
\end{equation}
where $\alpha$ and $\beta$ are the power-law index of photon energies below and above the break, respectively. $E_{p,obs}$ is the observed peak energy.

\begin{figure*}[ht!]\
\center
\resizebox{80mm}{!}{\includegraphics[]{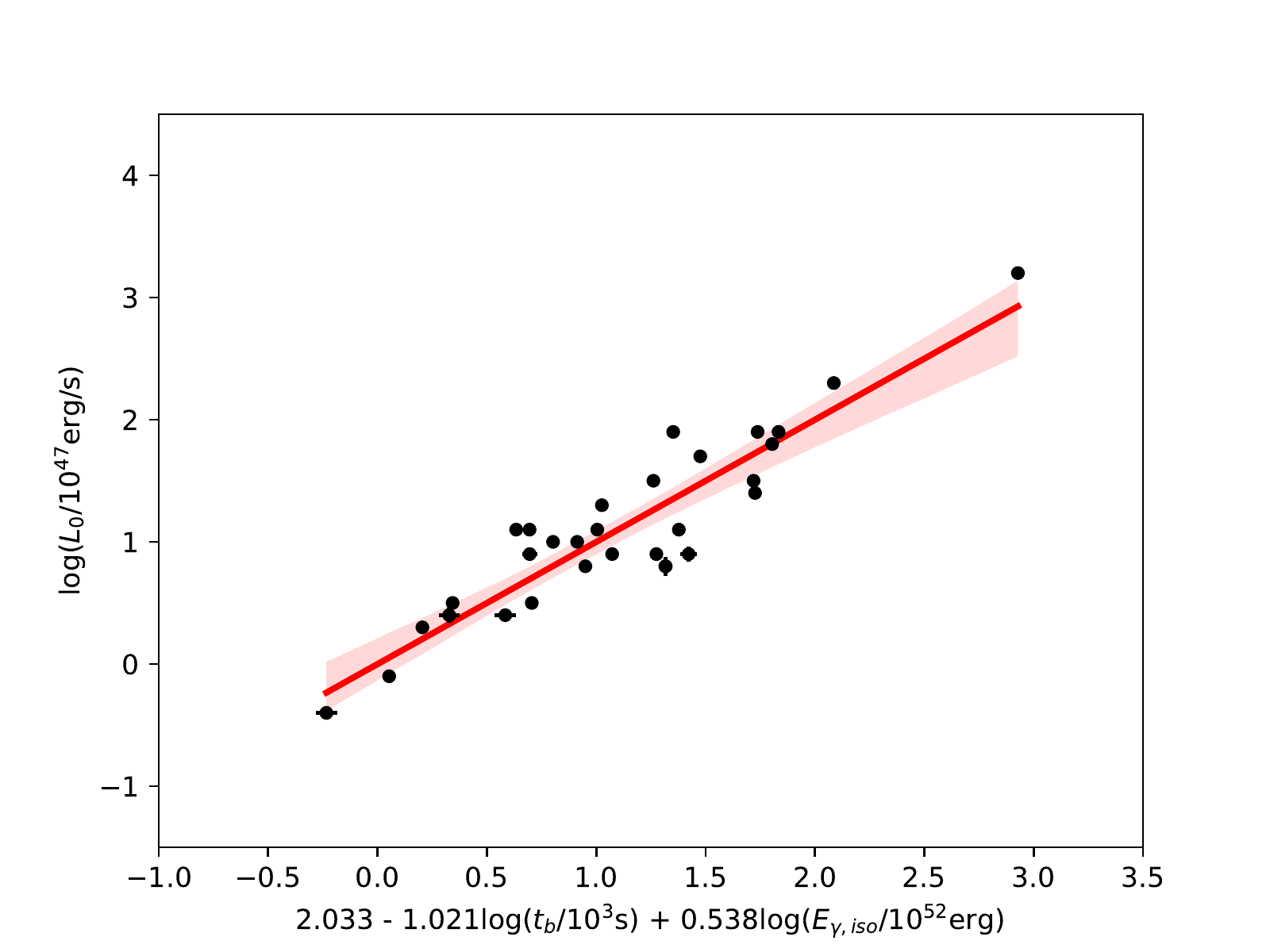}}\resizebox{80mm}{!}{\includegraphics[]{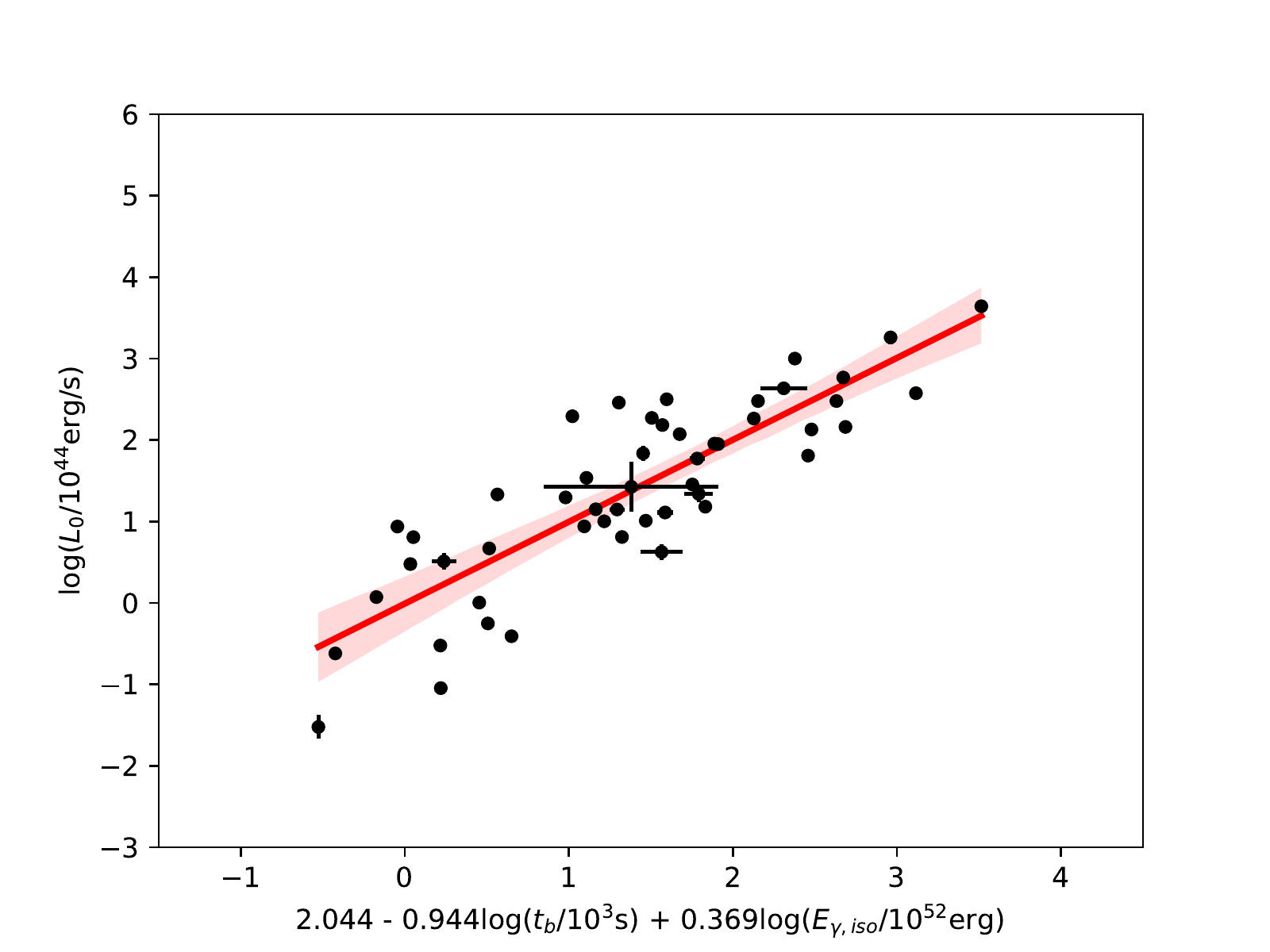}}\\
\caption{The correlation between luminosity $L_0$, the end time $t_b$ and the isotropic energy release $E_{\gamma,iso}$
($L_0 - t_b - E_{\gamma,iso}$).
Here we have set ${\Omega_m} = 0.3$ and $H_0 = 70~{\rm km~s^{-1}~Mpc^{-1}}$ for calculating the luminosity
from the measured flux. The data points are the GRBs in X-ray (left) and optical samples (right).
The red line corresponds to the best fitting values of the data points with a $95\%$ confidence band.}
\label{fig:1}
\end{figure*}

The best fitting results of a, b, c, and the intrinsic scatter $\sigma_{int}$ can be obtained by
using the likelihood function.
The corresponding likelihood can be written as~(\citealp{2005physics..11182D})

\begin{equation}\label{}
\begin{split}
 \mathcal{L}(a,b,c,\sigma_{int})&\propto\prod_i\frac{1}{\sqrt{\sigma_{int}^2 + \sigma_{y_i}^2 + b^2\sigma_{x_{1,i}}^2 + c^2\sigma_{x_{2,i}}^2}} \\
 &\times {\rm exp}[-\frac{{y_i - a - bx_{1,i} - cx_{2,i}}^2}{2(\sigma_{int}^2 + \sigma_{y_i}^2 + b^2\sigma_{x_{1,i}}^2 + c^2\sigma_{x_{2,i}}^2)}],
\end{split}
\end{equation}
where $\sigma_{int}$ is the extrinsic parameter. Here we have set $x_1 = {\rm log}(t_b/10^3s)$, $x_2 = {\rm log}
(E_{\gamma,iso}/10^{47}\rm erg)$
and $y = {\rm log}(L_0/10^{47}{\rm erg~s^{-1}})$ (for optical sample $y = {\rm log}(L_0/10^{44}{\rm erg~s^{-1}}$).
The minimization is performed employing the Markov Chain Monte Carlo (MCMC) algorithm with the \texttt{emcee} package~(\citealp{2013PASP..125..306F}).
Using X-ray afterglow sample with the platform stage, we find a tighter three-parameter correlation compared with
the previous two-parameter correlation. Figure \ref{fig:1} shows the correlation of equation \ref{eq7} for the X-ray
and optical samples. The best fitting results for the X-ray sample shown on the left panel are
$a = 2.033 \pm 0.144$, $b = -1.021 \pm 0.092$, $c = 0.538 \pm 0.144$ and $\sigma_{int} =0.303 \pm 0.043$,
corresponding to $L_{0} \propto t_{b}^{-1.021 \pm 0.092}E_{\gamma,iso}^{0.538 \pm 0.144}$.
The figure on the right shows that the best fitting results for optical sample are
$a = 2.044 \pm 0.149$, $b = -0.944 \pm 0.098$, $c = 0.369 \pm 0.088$ and
$\sigma_{int} = 0.614 \pm 0.066$, corresponding to $L_0 \propto t_b^{-0.944 \pm 0.098}E_{\gamma,iso}^{0.369 \pm 0.088}$.
The fitting results of the parameters are shown in Table~\ref{tab:corr}.

\begin{table*}[htbp]
  \centering
  \caption{The best fitting results of the parameters for different correlations}
    \begin{tabular}{c|c|c|c|cr}
  \hline
  \hline
    $L_0 - t_b - E_{\gamma,iso}$ correlation   & a & b & c & $\sigma_{\rm int}$ \\
  \hline
  X-ray sample & $2.033 \pm 0.144$ & $-1.021 \pm 0.092$ & $0.538 \pm 0.144$ & $0.303 \pm 0.043$\\
  Optical sample & $2.044 \pm 0.149$ & $-0.944 \pm 0.098$ & $0.369 \pm 0.088$ &
$0.614 \pm 0.066$ \\
  \hline
   Calibrated $L_0 - t_b - E_{\gamma,iso}$ correlation   & a & b & c & $\sigma_{\rm int}$ \\
  \hline
  X-ray sample & $1.948 \pm 0.203$& $-1.047 \pm 0.122$& $0.426 \pm 0.174$ & $0.317 \pm 0.078$\\
  Optical sample & $1.958 \pm 0.153$& $-1.027 \pm 0.106$& $0.423 \pm 0.088$ & $0.558 \pm 0.070$\\
  \hline
  $L_0 - t_b - E_{p,i}$ correlation   & $a^{\prime}$ & $b^{\prime}$ & $c^{\prime}$ & $\sigma_{\rm int}$ \\
  \hline
  X-ray sample & $1.556 \pm 0.487$& $-0.963 \pm 0.117$&$0.266 \pm 0.163$ & $0.363 \pm 0.054$\\
  Optical sample & $0.860 \pm 0.445$& $-0.900 \pm 0.104$& $0.604 \pm 0.172$ & $0.637 \pm 0.070$\\
  \hline
   Calibrated $L_0 - t_b - E_{p,i}$ correlation   & $a^{\prime}$ & $b^{\prime}$ & $c^{\prime}$ & $\sigma_{\rm int}$ \\
  \hline
  X-ray sample &$1.833 \pm 0.852$& $-1.008 \pm 0.154$& $0.126 \pm 0.333$ & $0.404 \pm 0.095$\\
  Optical sample &$0.741 \pm 0.480$& $-1.012 \pm 0.121$& $0.649 \pm 0.190$ & $0.616 \pm 0.080$\\
  \hline
  \hline
    \end{tabular}%
  \label{tab:corr}%
\end{table*}%

\subsection{Calibrating $L_0 - t_b - E_{\gamma,iso}$ correlation}

In above section, due to the lack of GRB data at low-redshift, we have fixed the values of ${\Omega_m}$ and $H_0$
to calculate the luminosity distance.
Therefore, one have to deal with the "circularity problem" when using the correlations of GRBs to constrain cosmological parameters.  Many methods have been developed to calibrate the correlation of GRBs in order to tackle this issue (\citealp{2008A&A...490...31C, 2008MNRAS.391L...1K, 2008ApJ...685..354L, 2011A&A...536A..96W, 2016A&A...585A..68W, 2019MNRAS.486L..46A}).
Here we will use the gaussian process (GP) method to calibrate the Dainotti Relation~(\citealp{2021MNRAS.507..730H, 2022ApJ...924...97W}).
This method is based on the fact that objects with the same redshift have the same luminosity distance in any model of the Universe.
In this work, GP regression is implemented using the public code GaPP~(\citealp{2012JCAP...06..036S}).
The calibrated correlation obtained in this method is model-independent. The process of calibrating correlation is divided into two main steps. First, ${\rm H}(z)$ data (\citealp{2018ApJ...856....3Y}) are used to calibrate the luminosity distance $d_L$ of GRBs with low redshift. According to the equations (\ref{eq:L0}) and (\ref{Eiso}), the newly calculated $E_{\gamma,iso}$ and $L_0$ are obtained,
and then the best fitting values of the $L_0 - t_b - E_{\gamma,iso}$ correlation parameters
for the calibrated low redshift GRBs are obtained. Second, the model-independent distance modulus of higher redshifts
are calculated using the calibrated parameters of fitting results of $L_0 - t_b - E_{\gamma,iso}$ correlation
in the lower redshifts.

We perform the GP process to reconstruct the continuous function $H(z)$, and one can refer to, e.g.,~\cite{2012JCAP...06..036S} and~\cite{2022ApJ...924...97W}
for more detailed discussions. Using the GP method, one can get the luminosity distance with respect to the Hubble parameter $H(z)$,

\begin{equation}\label{eq:dL_Hz}
d_L(z) = c(1+z){\int_0^z}\frac{dz}{H(z)}.
\end{equation}

\begin{figure*}[ht!]\
\center
\resizebox{95mm}{!}{\includegraphics[]{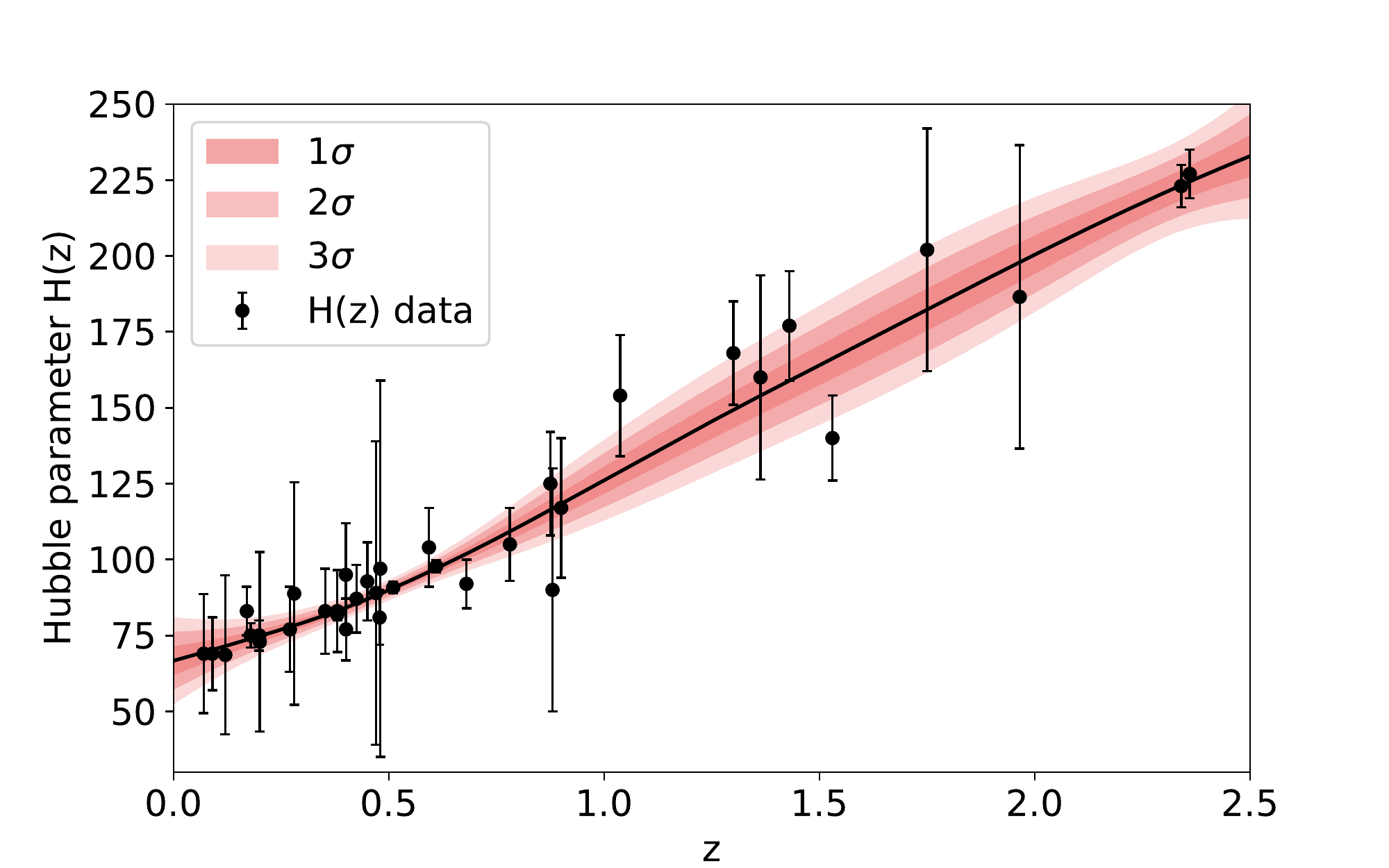}}\\
\caption{Reconstruction results of $H(z)$. The black line is the smoothed $H(z)$ function with GP method. The shaded regions correspond to $1\sigma$, $2\sigma$ and $3\sigma$ errors. The data points are 36 $H(z)$ data in the redshift range of $0.07\leq z\leq 2.36$. (Source: Figure 4 in~\cite{2021MNRAS.507..730H}.)}
\label{fig:Hz}
\end{figure*}

\begin{figure*}[ht!]\
\center
\resizebox{80mm}{!}{\includegraphics[]{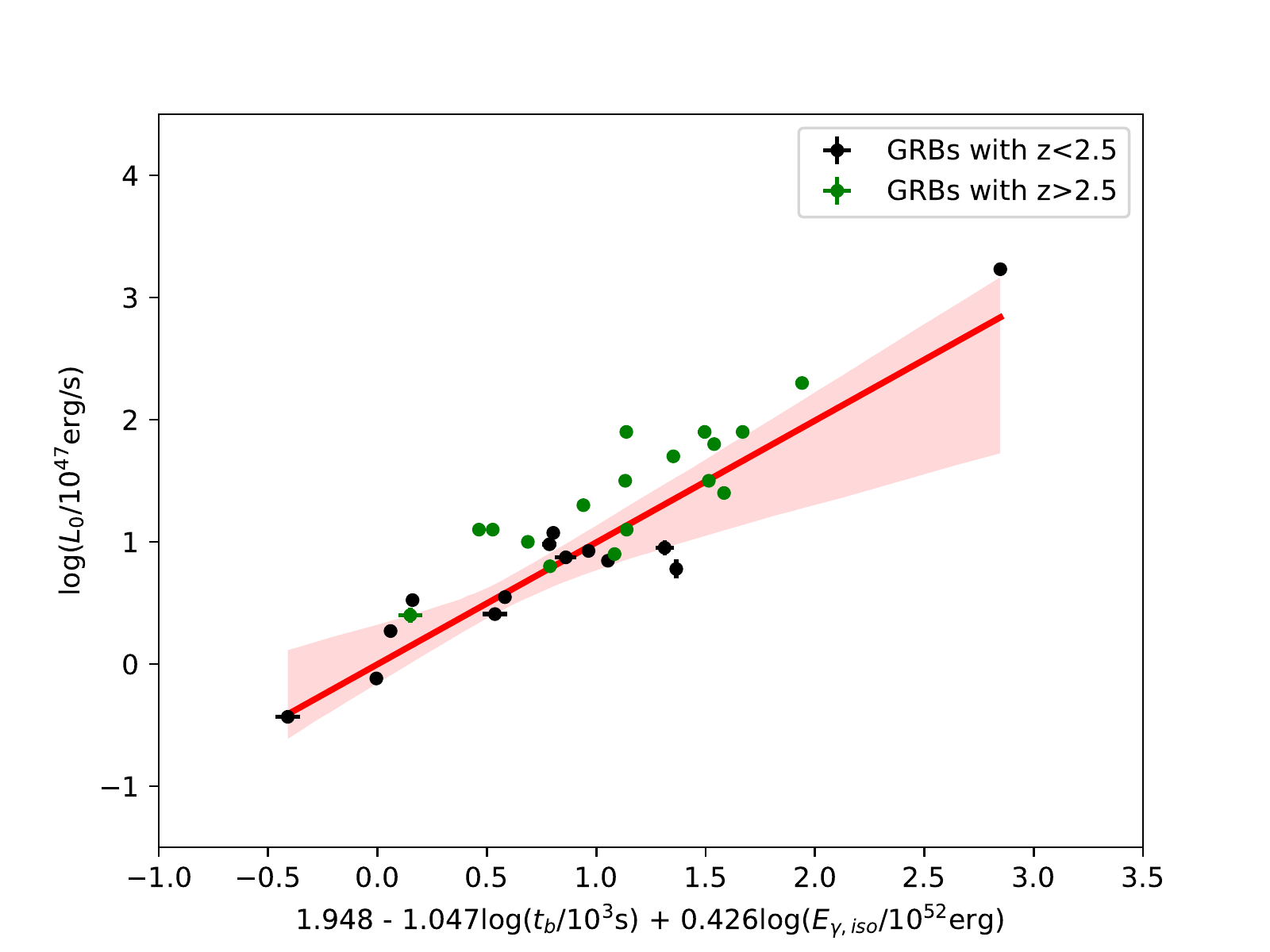}}\resizebox{80mm}{!}{\includegraphics[]{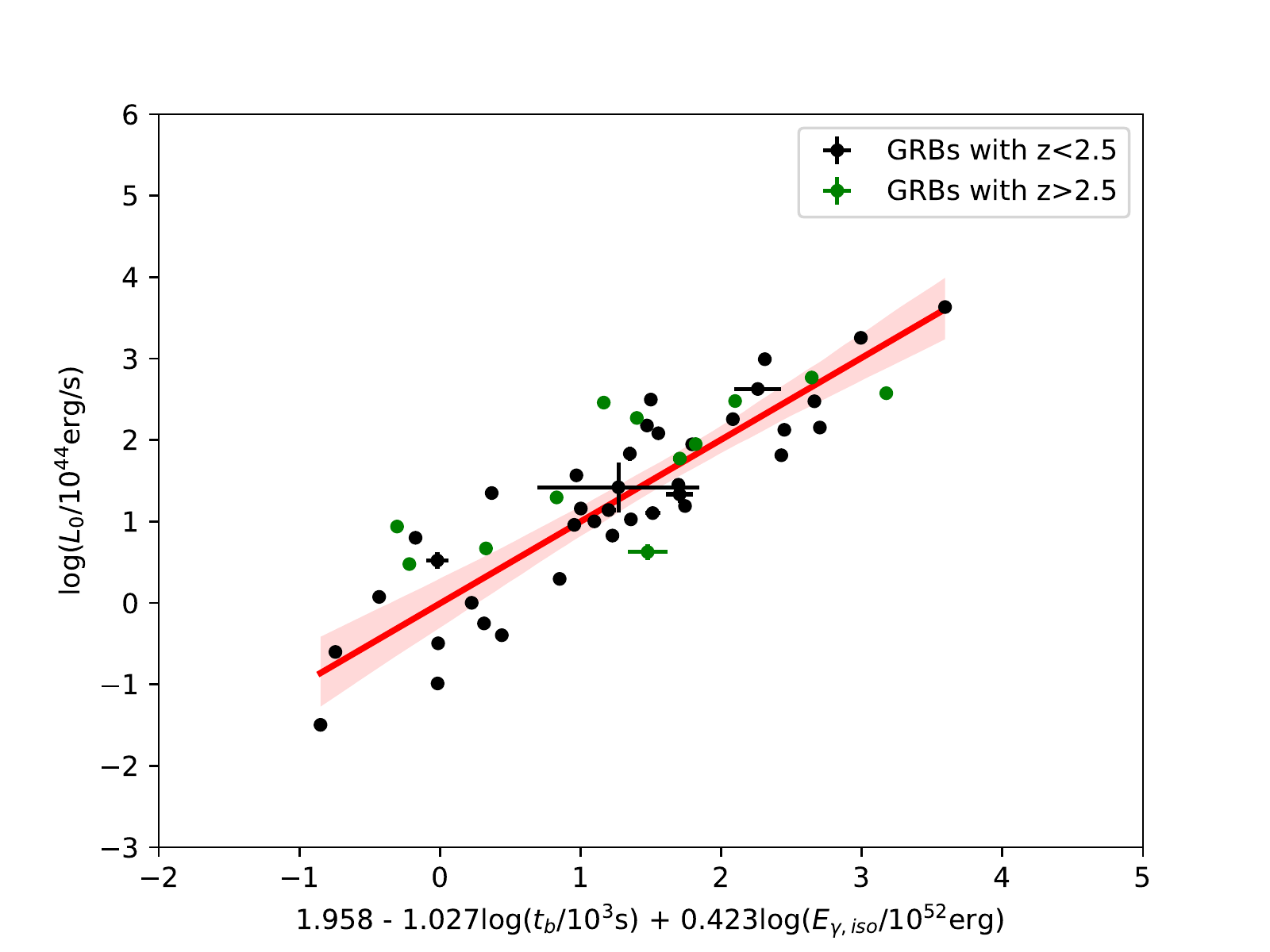}}
\caption{The calibrated $L_0 - t_b - E_{\gamma,iso}$ correlation for 14 X-ray sample (left) and
38 optical sample (right). GRBs with z\textgreater2.5 are also shown (green dots).
The red line is the best fitting line for data points, with a $95\%$ confidence band.}
\label{fig:2_calib}
\end{figure*}

After obtaining the continuous function, the values of $H(z)$ at different redshifts can be calculated.
Fig.~\ref{fig:Hz} shows the reconstruction results of the $H(z)$ curve. According to the equation (\ref{eq:dL_Hz}),
we can get the corresponding redshift luminosity distance of each GRB, and then we get the $L_0$ and $E_{\gamma,iso}$
from the equations (\ref{eq:L0}) and (\ref{Eiso}), which can be used to fit the parameters $a$, $b$ and $c$
of the correlation $L_{0} - t_{b} - E_{\gamma,iso}$. We have used 36 $H(z)$ data collected
by~\cite{2018ApJ...856....3Y} in the redshift range of $0.17<z<2.36$ for the calibration process.
Therefore, we can estimate the luminosity distance of GRBs in the redshift of $z\lesssim 2.50$.
There are 14 GRBs of X-ray sample and 38 GRBs of optical sample in this
redshift range, and then we use the two selected samples to obtain the model-independent luminosity distances,
which can be used to calibrate the $L_0 - t_b - E_{\gamma,iso}$ correlation for the X-ray and optical samples, respectively.
The corresponding results after calibration are shown in Figure \ref{fig:2_calib}.
The best fitting results of the selected X-ray sample are $a = 1.948 \pm 0.203$, $b = -1.047 \pm 0.122$, $c = 0.426 \pm 0.174$ and $\sigma_{\rm int} =0.317 \pm 0.078$, corresponding to $L_0 \propto t_b^{-1.047 \pm 0.122}E_{\gamma,iso}^{0.426 \pm 0.174}$.
For the selected optical sample, the best fitting results are $a = 1.958 \pm 0.153$, $b = -1.027 \pm 0.106$, $c = 0.423 \pm 0.088$ and $\sigma_{\rm int} =0.558 \pm 0.070$, corresponding to $L_0 \propto t_b^{-1.027 \pm 0.106}E_{\gamma,iso}^{0.423 \pm 0.088}$.
The fitting results of the parameters are shown in Table~\ref{tab:corr}.

The $L_0 - t_b - E_{\gamma,iso}$ correlation calibrated in this way are model-independent and can be used to constrain cosmological parameters.
For the optical sample, the correlation after calibration is tighter than that obtained by fitting the entire samples.
Furthermore, for X-ray sample, the value of $\sigma_{\rm int}$ after calibration is slightly larger than that
obtained for all samples. The reason could be that there are not enough data points for X-ray samples in lower redshifts,
leading to the increasing of the internal dispersion.

In the process of constraining cosmological parameters by employing the calibrated correlation $L_0 - t_b - E_{\gamma,iso}$,
we found that it is difficult to get robust constraints on cosmological parameters.
The reasons of why $L_0 - t_b - E_{\gamma,iso}$ correlation is not good cosmological probe has been discussed in~\cite{2021ApJ...920..135X}, and our results support that states.
First, the method of extrapolating the calibration results from lower redshifts to higher redshifts may not be appropriate,
due to the possible evolution of the $L_0 - t_b - E_{\gamma,iso}$ correlation.
Second, the selected GRB samples that may have the same physical mechanism are not enough to be
calibrated for lower redshifts, leading to the increase of the internal dispersion.
Third, the calculations of both $L_0$ and $E_{\gamma,iso}$ depend on the luminosity distance $d_L$, which relies on the cosmological parameters. This means that cosmological effect may be largely cancelled out in the relationship,
or the correlation is insensitive to the cosmological parameters.

\subsection{Calibrating $L_0 - t_b - E_{p,i}$ correlation}

We now investigate whether the $L_0-t_b-E_{p,i}$ correlation can be used to probe the cosmological parameters,
where $E_{p,i} = E_{p,obs}\times(1+z)$ is the spectral peak energy with the observed peak energy $E_{p,obs}$.
For this correlation, we have collected data from the following literature :~\cite{2018ApJ...863...50S}, \cite{2020MNRAS.492.1919M}, \cite{2021ApJ...920..135X} and \cite{2021MNRAS.508...52L},
and we have removed GRB190114A due to the lack of $E_{p,i}$ data.

We studied the $L_0 - t_b - E_{p,i}$ correlation using the same approach described in the previous section.
The relation can be written as

\begin{equation}\label{Ep_relation}
{\rm log}\frac{L_0}{10^{47}{\rm erg~s^{-1}}} = a^{\prime} + b^{\prime}{\rm log}\frac{t_b}{10^3s} + c^{\prime}{\rm log}\frac{E_{p,i}}{\rm keV}.
\end{equation}

\begin{figure*}[ht!]\
\center
\resizebox{80mm}{!}{\includegraphics[]{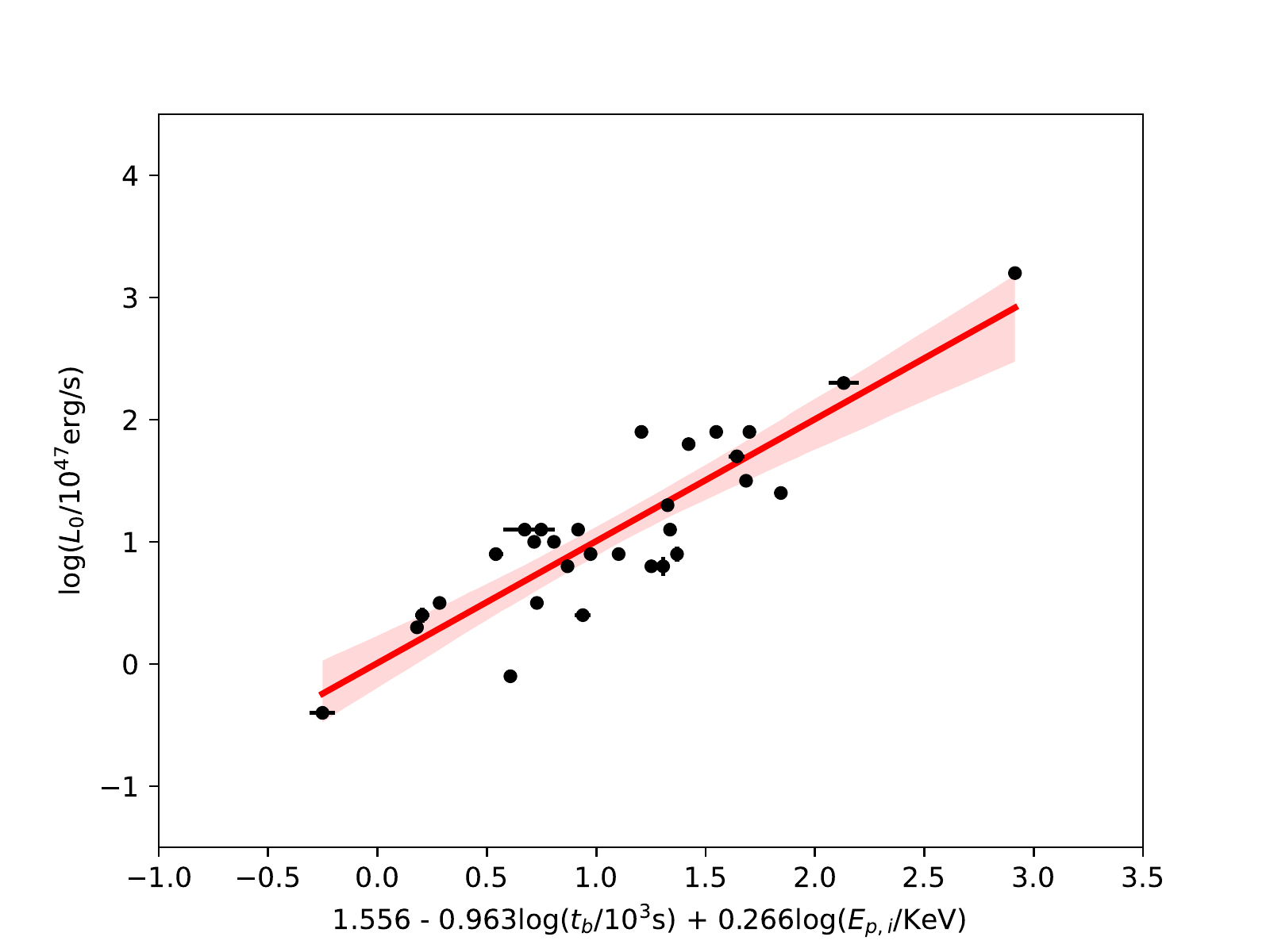}}\resizebox{80mm}{!}{\includegraphics[]{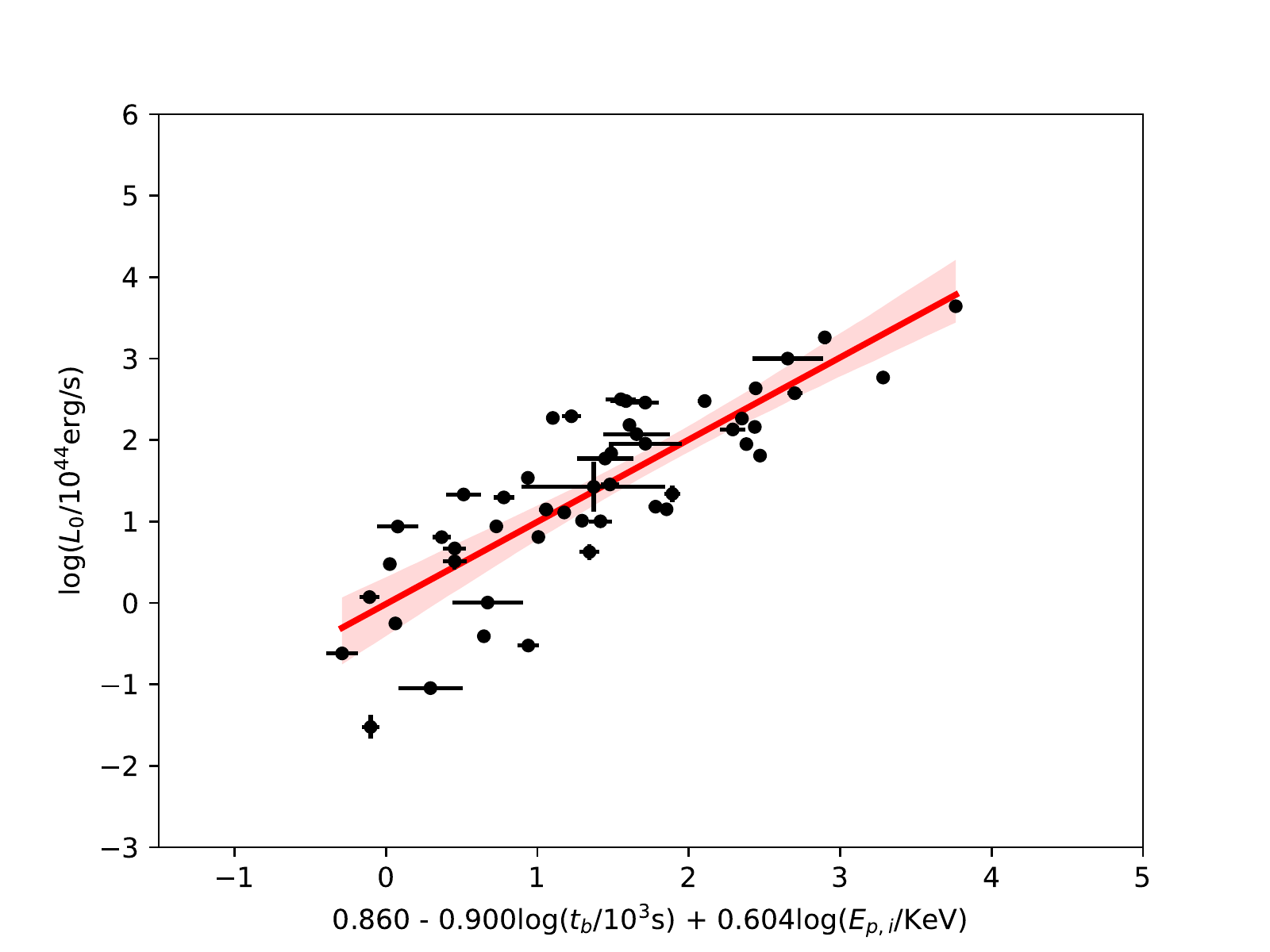}}
\caption{The correlation between luminosity $L_0$, the end time $t_b$ and the spectral peak energy $E_{p,i}$ ($L_0 - t_b - E_{p,i}$).
The data points are the GRBs in X-ray (left) and optical samples (right). The red line corresponds to the best fitting values of the data points, with a $95\%$ confidence band.}
\label{fig:Fitting_Ep}
\end{figure*}

For the X-ray sample, the best fitting results are
$a^{\prime} = 1.556 \pm 0.487$, $b^{\prime} = -0.963 \pm 0.117$, $c^{\prime} = 0.266 \pm 0.163$ and $\sigma_{\rm int} =0.363 \pm 0.054$,
corresponding to $L_0 \propto t_b^{-0.977 \pm 0.111}E_{p,i}^{0.240 \pm 0.141}$. For the optical sample,
the best fitting result are $a^{\prime} = 0.860 \pm 0.445$, $b^{\prime} = -0.900 \pm 0.104$, $c^{\prime} = 0.604 \pm 0.172$ and $\sigma_{\rm int} =0.637 \pm 0.070$, corresponding to $L_0 \propto t_b^{-0.900 \pm 0.104}E_{p,i}^{0.604 \pm 0.172}$.
The corresponding fitting results are shown in Figure~\ref{fig:Fitting_Ep}.
It can be seen that there is a clear correlation
between these three parameters. The fitting results of the parameters are shown in Table~\ref{tab:corr}.

\begin{figure*}[ht!]\
\center
\resizebox{80mm}{!}{\includegraphics[]{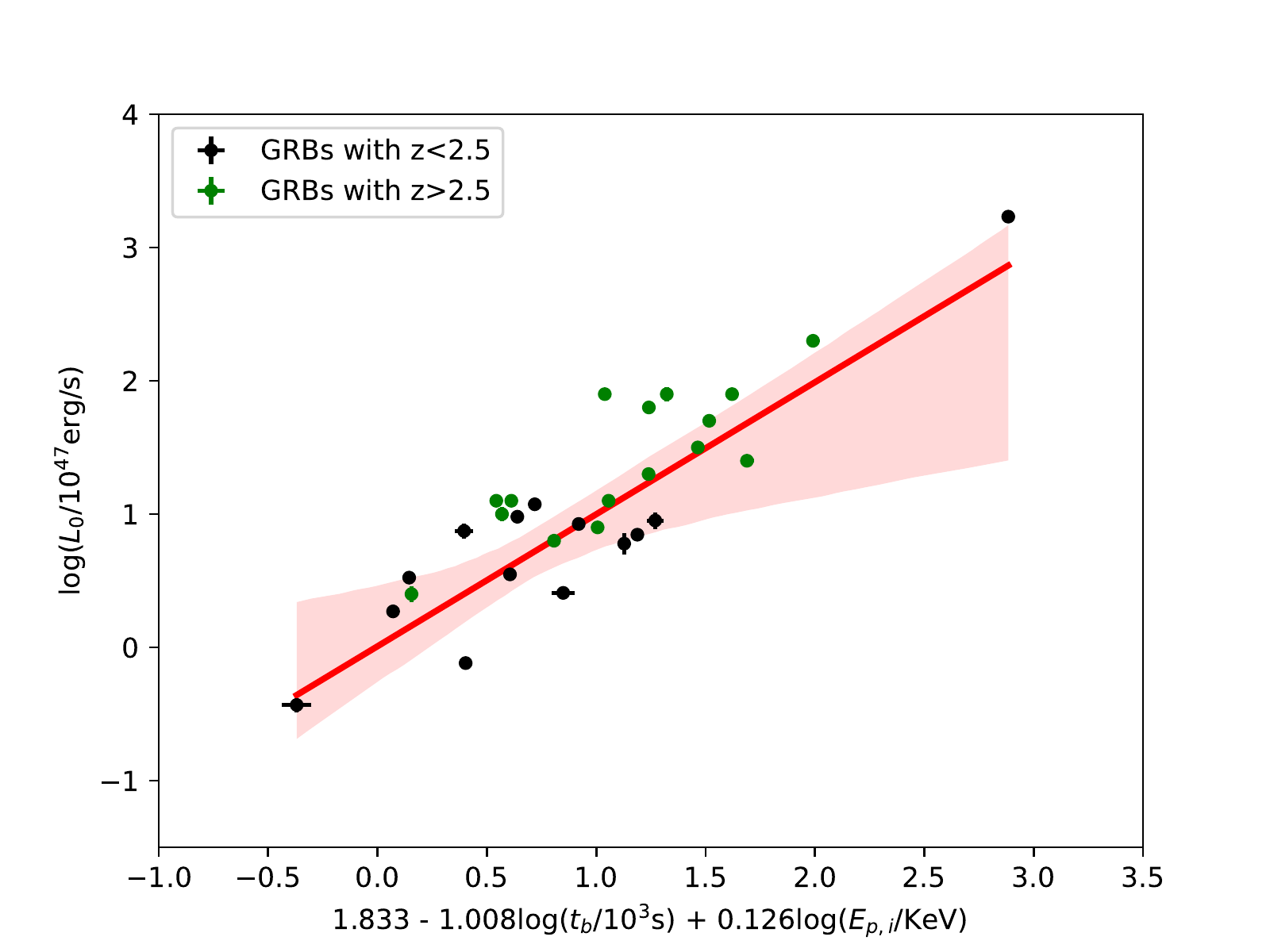}}\resizebox{80mm}{!}{\includegraphics[]{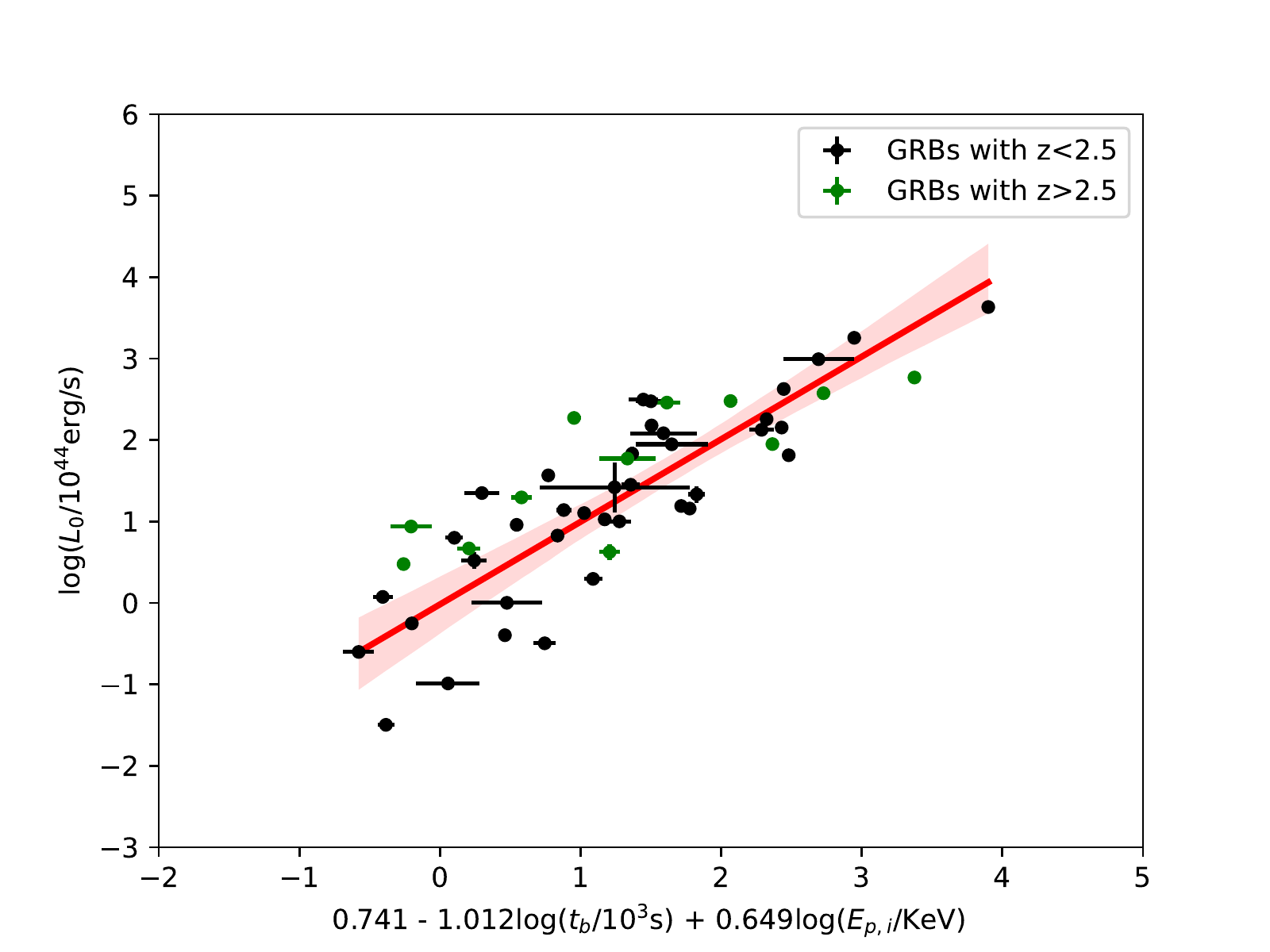}}
\caption{The calibrated $L_0 - t_b - E_{p,i}$ correlation for 14 X-ray sample (left) and 38 optical sample (right).
GRBs with z\textgreater2.5 are also plotted (green dots). The red line corresponds to the best fitting values
of the data points, with a $95\%$ confidence band.}
\label{fig:Fitting_Ep_calib}
\end{figure*}

\begin{figure*}[ht!]\
\center
\resizebox{80mm}{!}{\includegraphics[]{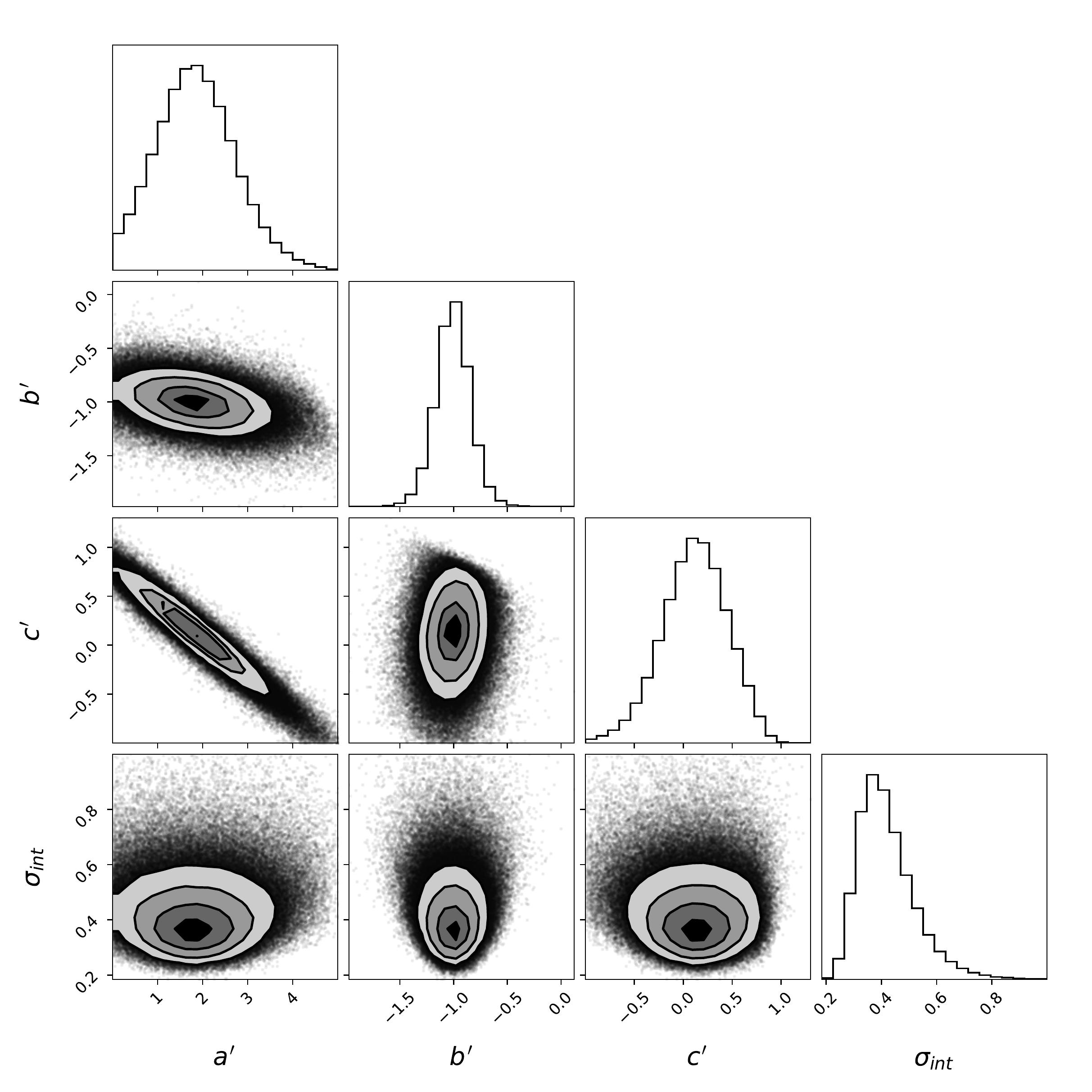}}\resizebox{80mm}{!}{\includegraphics[]{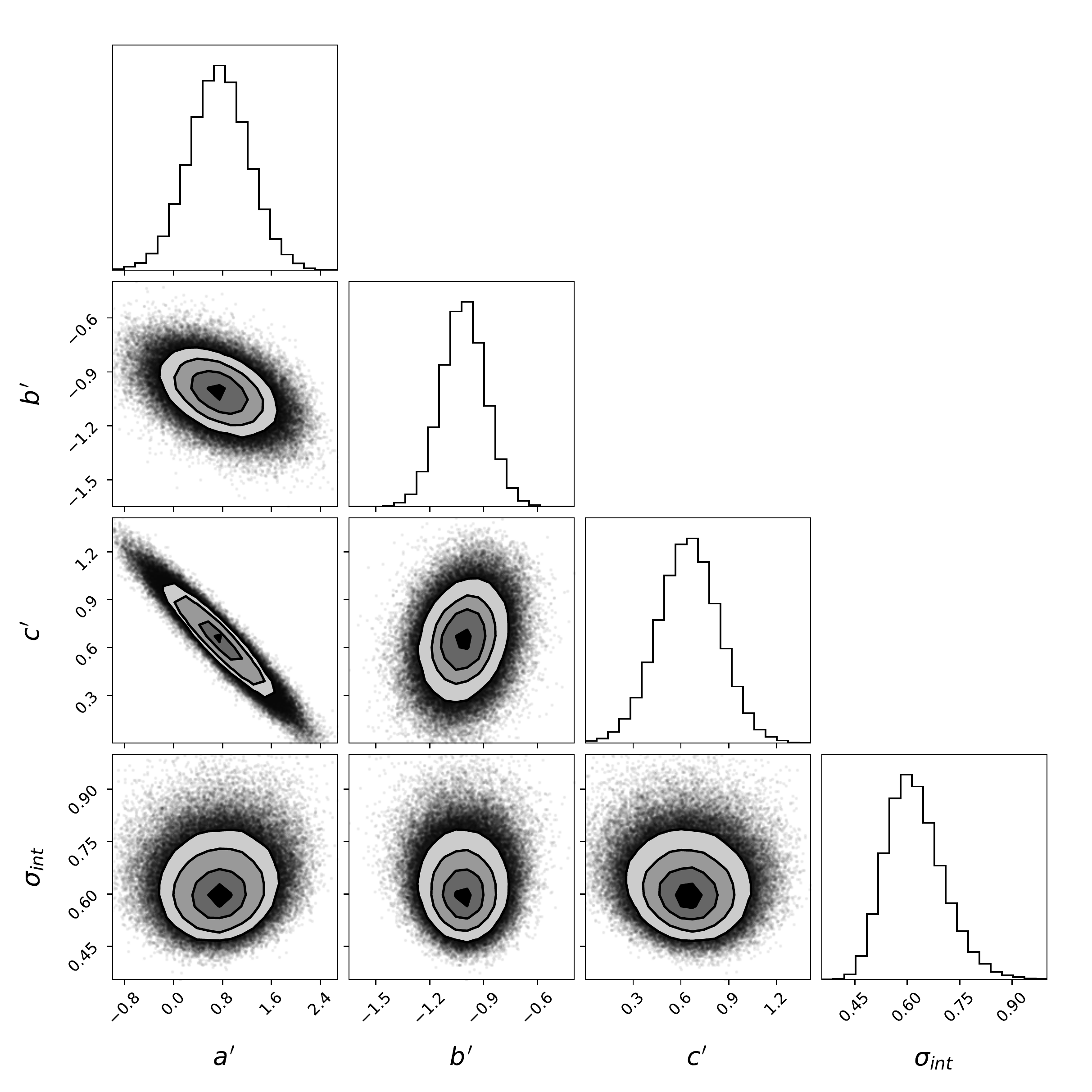}}
\caption{The corner diagrams of the related parameters used in Fig.~\ref{fig:Fitting_Ep_calib}
for the calibrated $L_0 - t_b - E_{p,i}$ correlation for the X-ray sample (left) and optical sample (right).
The shaded regions correspond to 1$\sigma$, 2$\sigma$ and 3$\sigma$ errors.}
\label{fig:Fitting_Ep_calib_cor}
\end{figure*}

We also calibrated the three-parameter correlation using the selected data in
the redshift range of $z<2.5$ for the two sets of samples. The corresponding results after calibration are
shown in Figure~\ref{fig:Fitting_Ep_calib}. For the X-ray sample, the best fitting result are $a^{\prime} = 1.833 \pm 0.852$, $b^{\prime} = -1.008 \pm 0.154$, $c^{\prime} = 0.126 \pm 0.333$ and $\sigma_{\rm int} =0.404 \pm 0.095$, corresponding to $L_0 \propto t_b^{-1.008 \pm 0.154}E_{p,i}^{0.126 \pm 0.333}$. The best fitting result for the low redshift optical sample
are $a^{\prime} = 0.741 \pm 0.480$, $b^{\prime} = -1.012 \pm 0.121$, $c^{\prime} = 0.649 \pm 0.190$ and $\sigma_{\rm int} =0.616 \pm 0.080$, corresponding to $L_0 \propto t_b^{-1.012 \pm 0.121}E_{p,i}^{0.649 \pm 0.190}$.
The corresponding corner diagrams can be found in Figure~\ref{fig:Fitting_Ep_calib_cor},
and the fitting results of the parameters are shown in Table~\ref{tab:corr}.
We will use the calibrated $L_0 - t_b - E_{p,i}$ correlation for cosmological constraints.

\section{constraints on the cosmological parameters using the three-parameter correlation}

In the flat $\Lambda$CDM model, the distance modulus can be written as

\begin{equation}\label{mu_th}
\mu_{{\rm th}} = 5{\rm log}\frac{d_L}{{\rm Mpc}} + 25 = 5{\rm log}\frac{d_L}{\rm cm} + 97.45.
\end{equation}

From the equations (\ref{eq:L0}) and (\ref{Ep_relation}) one can get the corresponding function to replace $d_L$
in the equation (\ref{mu_th}). When optical sample is considered,
the typical value of $L_0$ is $10^{44}\rm erg~s^{-1}$. The corresponding observed distance can be defined as

\begin{eqnarray}
\mu_{obs} = &&\frac{5}{2}\Bigg[a + b({\rm log}t_b-3) + c{\rm log}E_{p,i} \nonumber \\
            &&-{\rm log}\frac{4{\pi}F}{(1+z)} + 44\Bigg] -97.45.
\label{mu_obs}
\end{eqnarray}

The uncertainty of $\mu_{obs}$ can be expressed as

\begin{equation}\label{mu_obs_err}
\begin{split}
\sigma_{obs}& = \frac{5}{2}\Bigg[\sigma_{\rm int}^2 + \sigma_a^2 + \sigma_b^2({\rm log}t_b-3)^2 + b^2(\frac{\sigma_{t_b}}{t_b {\rm ln}10})^2 \\
&+ c^2(\frac{\sigma_{E_{p,i}}}{E_{p,i} {\rm ln}10})^2 + (\frac{\sigma_{F}}{F {\rm ln}10})^2 + \sigma_c^2{\rm log}^2_{E_{p,i}}\Bigg]^{1/2},
\end{split}
\end{equation}
where $\sigma_i$ represents the error of different parameters derived as $\sqrt{(\sigma_u^2 + \sigma_d^2)^2/2}$,
where $\sigma_u$ and $\sigma_d$ represent the upper and lower error of the parameters, respectively.
The best-fitting parameters can be obtained by minimizing the $\chi^2$,

\begin{equation}\label{chi^2}
\chi^2= \sum_{j=1}^{N}\frac{\left[\mu_{obs}(z) - \mu_{th}(\Omega_i,z)\right]^2}{\sigma_{obs}^2},
\end{equation}
where N is the number of samples in each category. $\mu_{obs}$ and $\sigma_{obs}$ can be calculated from
the equations (\ref{mu_obs}) and (\ref{mu_obs_err}), respectively.
$\mu_{th}(\Omega_i,z)$ is the theoretical distance modulus, and
$\Omega_i$ represent the cosmological parameters needed to be constrained.

In general, the luminosity distance can be written as

\begin{equation}\label{dl_nonflat}
d_L=\left\{
\begin{array}{lcl}
\frac{c(1+z)}{H_0}(-\Omega_k)^{-\frac{1}{2}}{\rm sin}\left[(-\Omega_k)^{-\frac{1}{2}}{\int_0^z}\frac{dz}{E(z)}\right]
& & {\Omega_k < 0}, \\
\frac{c(1+z)}{H_0}{\int_0^z}\frac{dz}{E(z)}& & {\Omega_k = 0},\\
\frac{c(1+z)}{H_0}\Omega_k^{-\frac{1}{2}}{\rm sinh}\left[\Omega_k^{-\frac{1}{2}}{\int_0^z}\frac{dz}{E(z)}\right] & & {\Omega_k > 0},
\end{array}
\right.
\end{equation}
where $\Omega_k$ stands for the curvature of space. $E(z)$ can be expressed as

\begin{equation}\label{eq15}
E(z) = \sqrt{\Omega_m(1+z)^3 + (1 - \Omega_m - \Omega_{\Lambda})(1+z)^2 + \Omega_{\Lambda}}.
\end{equation}

Now one can get the corresponding theoretical distance modulus $\mu_{th}$
by taking the equation (\ref{dl_nonflat}) into (\ref{mu_th}).
Using the minimized equation (\ref{chi^2}), one can obtain the constrains
on the related parameters of the $\Lambda$CDM cosmological model.

\begin{figure*}[ht!]\
\center
\resizebox{80mm}{!}{\includegraphics[]{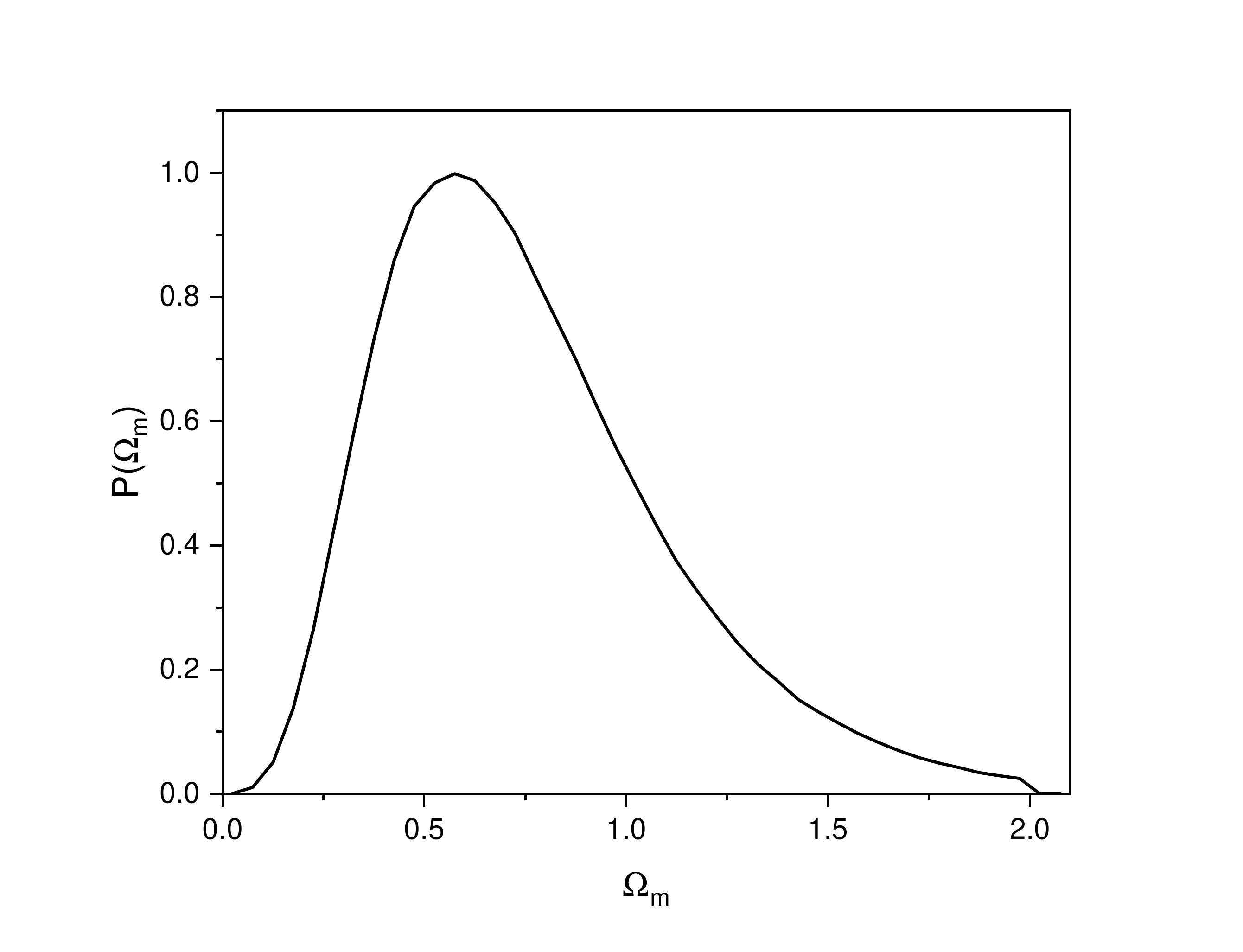}}\resizebox{70mm}{!}{\includegraphics[]{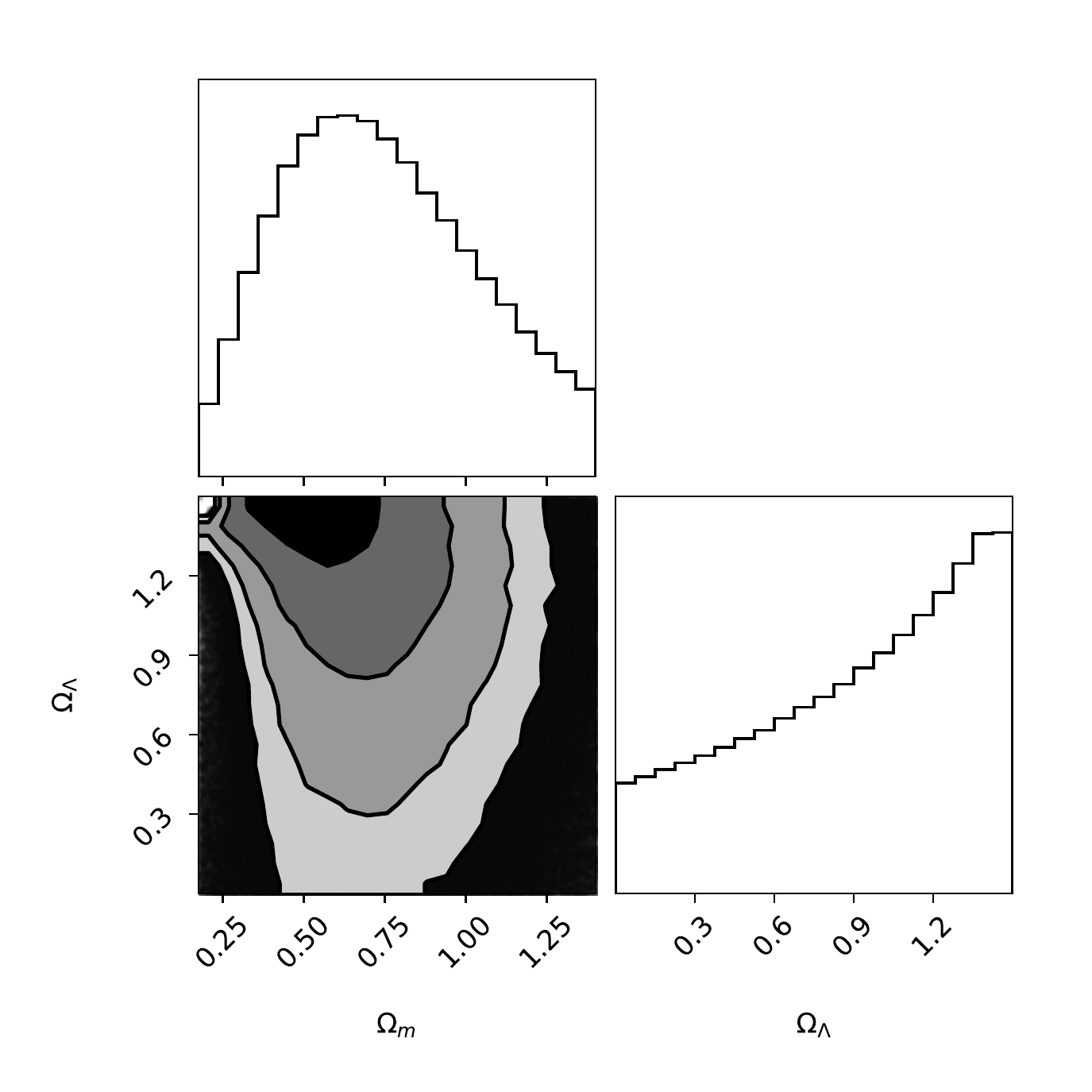}}\\
\caption{Constraints on the cosmological parameters in $\Lambda$CDM universe model using the optical sample.
The left panel shows the probability distribution of $\Omega_m$ in the flat $\Lambda$CDM model. The right panel shows the constraints on $\Omega_m$ and $\Omega_{\Lambda}$ in the non-flat $\Lambda$CDM model. }
\label{fig:Optical_limit}
\end{figure*}

\begin{figure*}[ht!]\
\center
\resizebox{80mm}{!}{\includegraphics[]{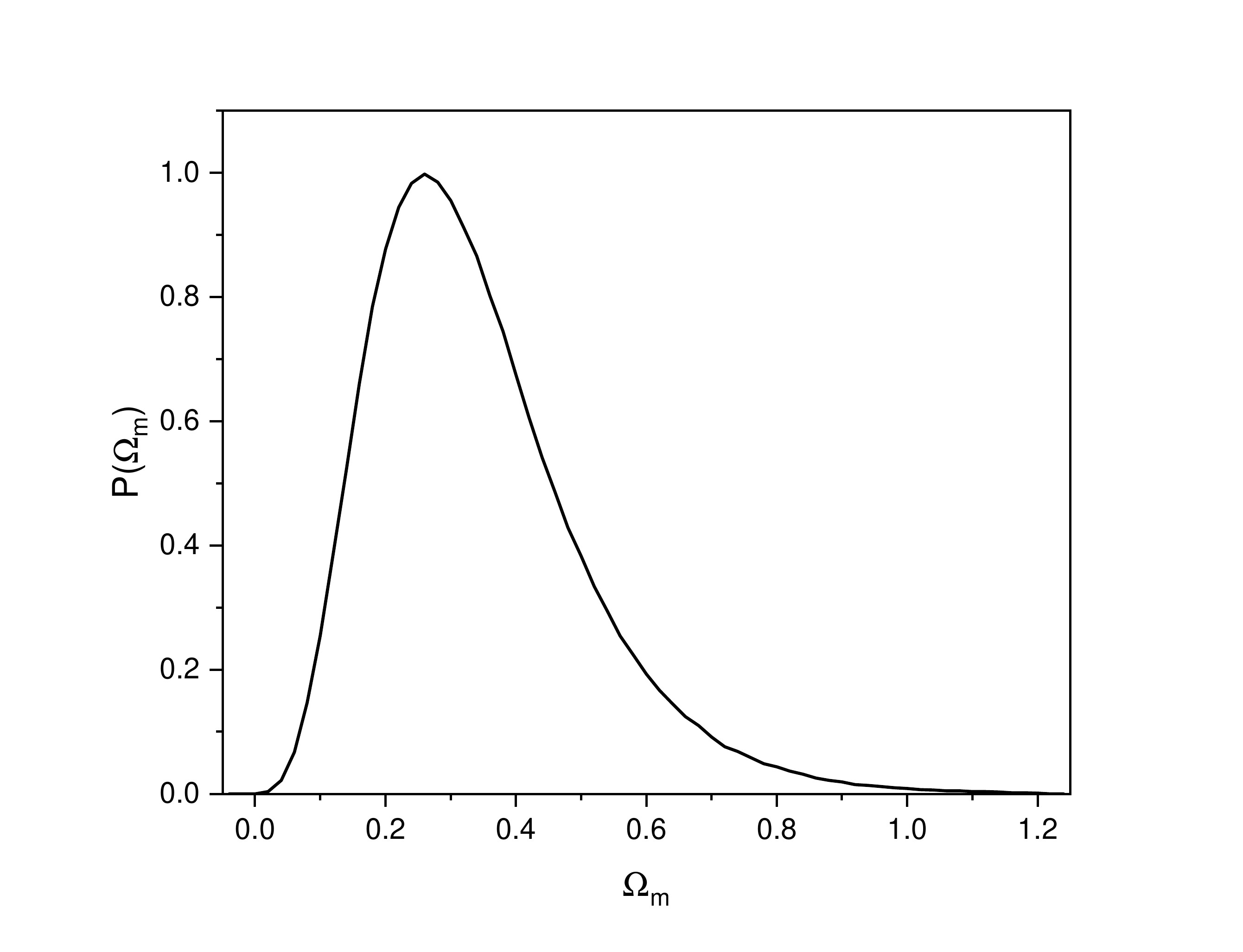}}\resizebox{70mm}{!}{\includegraphics[]{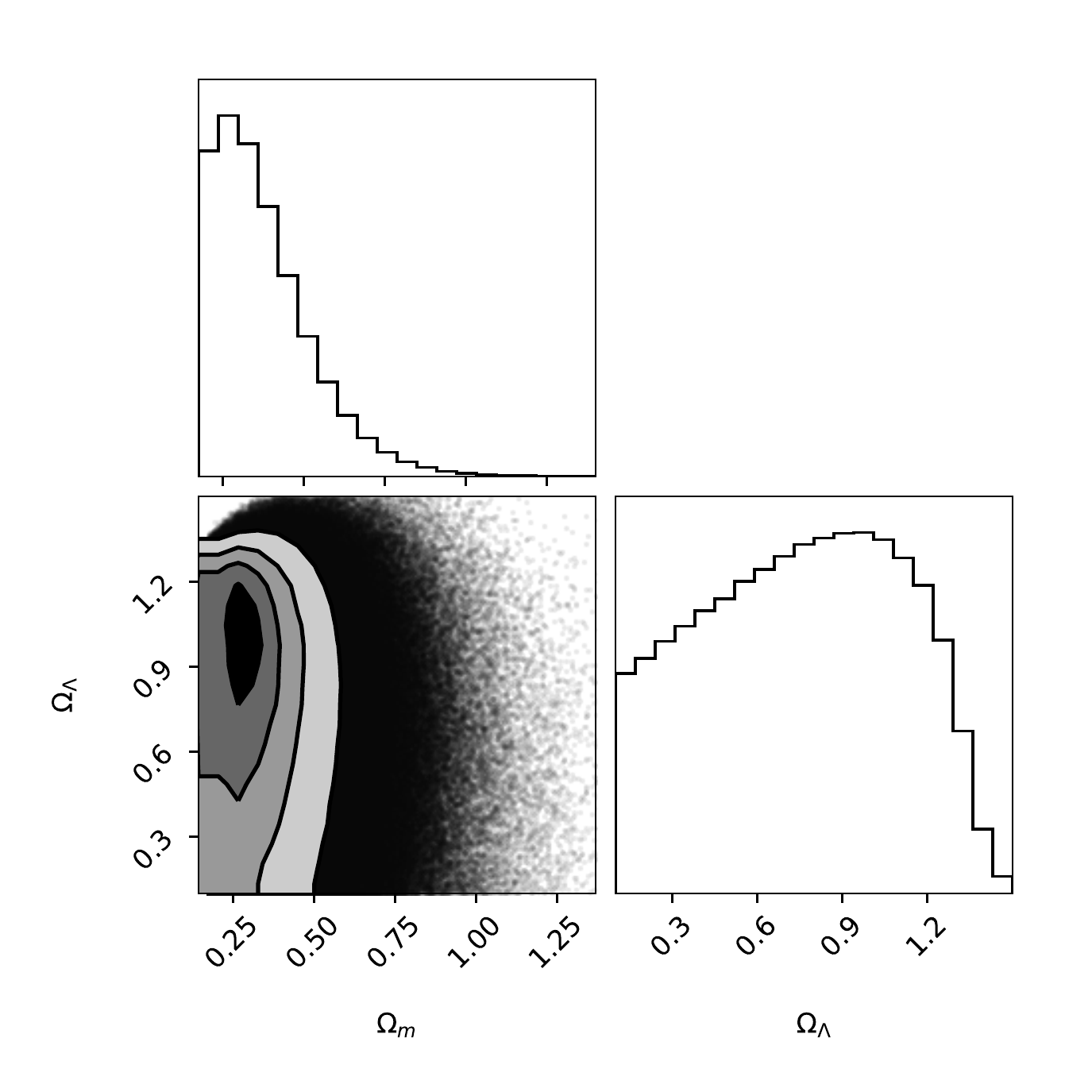}}\\
\caption{Constraints on the cosmological parameters in $\Lambda$CDM model using the optical and X-ray samples.
The left panel shows the probability distribution of $\Omega_m$ in the flat $\Lambda$CDM model. The right panel shows the constraints on $\Omega_m$ and $\Omega_{\Lambda}$ in the non-flat $\Lambda$CDM model.}
\label{fig:Optical+X-ray_limit}
\end{figure*}

Only employing the optical sample distributed in the redshift range of $0.13\lesssim z\lesssim 4.67$,
we get the best-fitting results of the cosmological parameter $\Omega_m = 0.697_{-0.278}^{+0.402}(1\sigma)$
for a flat universe model, as shown in the left panel of Fig.~\ref{fig:Optical_limit}. For the non-flat model,
the best-fitting results are $\Omega_m = 0.713_{-0.278}^{+0.346}$, $\Omega_{\Lambda} = 0.981_{-0.580}^{+0.379}(1\sigma)$, as shown in the right panel of Figure~\ref{fig:Optical_limit}. We only combine the optical and X-ray samples to constrain the parameters.
For a flat $\Lambda$CDM model, the left panel of Fig.~\ref{fig:Optical+X-ray_limit} shows the best-fit
matter density parameter $\Omega_m = 0.313_{-0.125}^{+0.179}(1\sigma)$.
The best-fitting results $\Omega_m = 0.344_{-0.112}^{+0.176}$, $\Omega_{\Lambda} = 0.770_{-0.416}^{+0.366}(1\sigma)$
for a non-flat $\Lambda$CDM model are shown in the right of Figure~\ref{fig:Optical+X-ray_limit}.
The final constraints on the cosmological parameters can be found in Table~\ref{tab:cos_para}.
It is worth noting that the combined samples reduce the uncertainties of the parameters compared to the optical GRB sample alone.
Note that the constrains on the cosmological parameters obtained from the X-ray and optical samples of GRBs are
weaker than that obtained from SNe Ia and X-ray sample only of~\cite{2022ApJ...924...97W}. There are several reasons for this.
One is that the optical sample size is not large enough. The second is that the optical sample might be not clean.
Compared with X-ray, optical band has lower radiation efficiency and more complex radiation composition.
In the future, with more powerful telescopes to obtain multi-band spectra, we should be able to reveal the intrinsic mechanism. It is also hoped that the optical samples can be classified more carefully by the spectral index and light-curve shapes. The third point is based on the fact that there are measurement inaccuracies in the three parameters of $L_0$, $t_b$, and $E_{p,i}$~(\citealp{2018ApJ...863...50S, 2010ApJ...725.2209L, 2012ApJ...758...27L,2020ApJ...900..112Z}).

\begin{figure*}[ht!]\
\center
\resizebox{80mm}{!}{\includegraphics[]{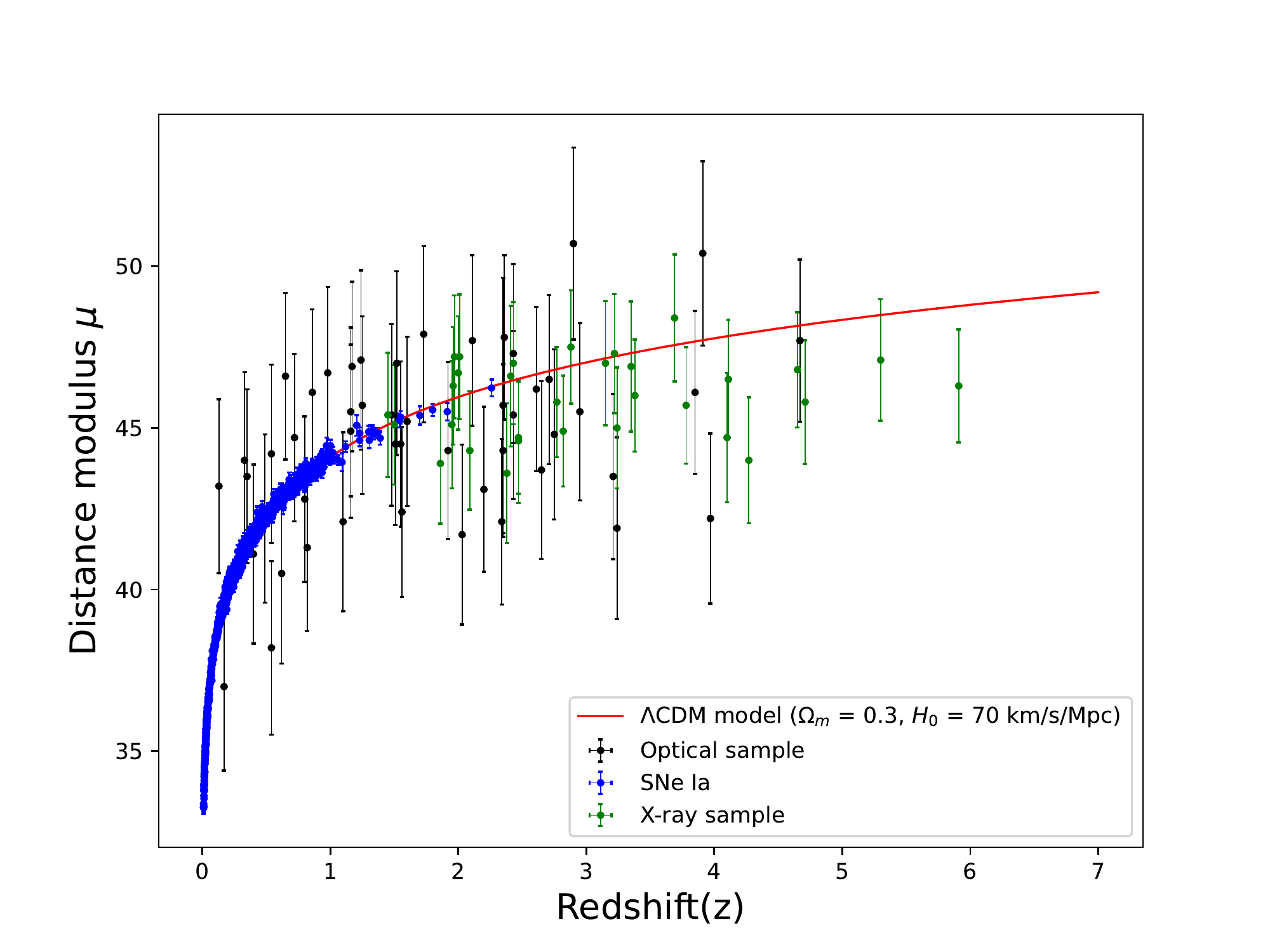}}\resizebox{80mm}{!}{\includegraphics[]{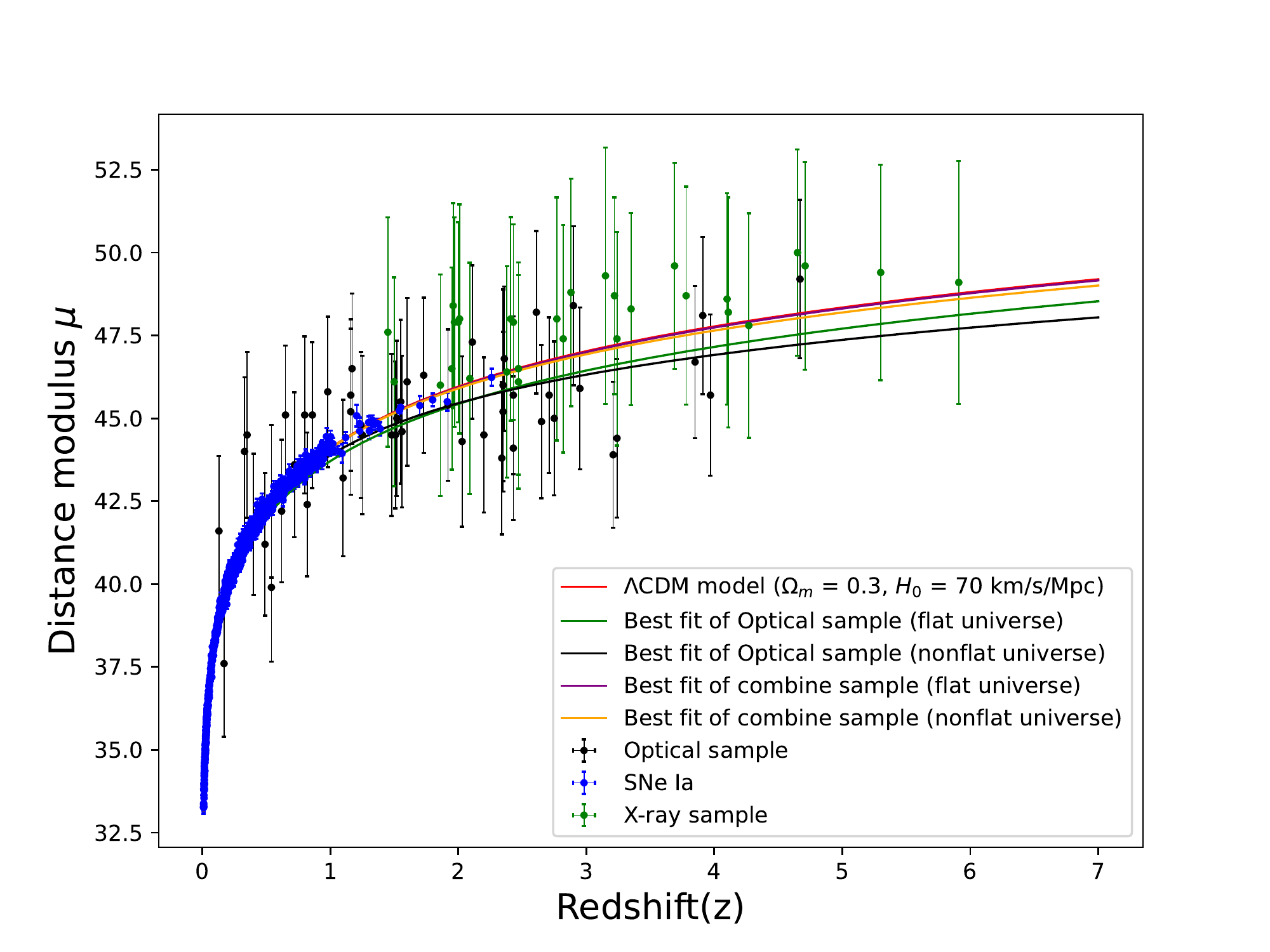}}
\caption{Calibrated GRB Hubble diagram using the $L_0 - t_b - E_{\gamma,iso}$ correlation (left) and $L_0 - t_b - E_{p,i}$ correlation (right). The black and green points represent optical and X-ray samples, respectively. Blue points are SNe Ia from the Pantheon sample.
The solid red line corresponds to the theoretical distance modulus calculated for a flat $\Lambda$CDM model
with $H_0 = 70~\rm km~s^{-1}~Mpc^{-1}$ and $\Omega_m = 0.3$. For a flat universe, the best fit from the optical sample is shown as a green line and the best fit from the combine sample is shown as a purple line. For the nonflat universe, the black line shows the best fit from the optical sample, and the orange line shows the best fit from the combine sample.}
\label{fig:Hubble diagram}
\end{figure*}

Additionally, the Hubble diagram derived from the $L_0 - t_b - E_{\gamma,iso}$ correlation for all the calibrated samples is shown in the left
panel of Figure~\ref{fig:Hubble diagram}.
It can be seen from the diagram that the points at high-redshift of all the samples are highly dispersive and accompanied by large error bars, making it difficult obtain meaningful constraints on the cosmological parameters with this correlation.
In the right of Figure~\ref{fig:Hubble diagram}, we present the calibrated GRB Hubble diagram from the $L_0 - t_b - E_{p,i}$ correlation.
It can be seen that the dispersion of the data points in the Hubble diagram from the $L_0 - t_b - E_{p,i}$ correlation is smaller than that in the $L_0-t_b-E_{\gamma,iso}$ correlation. It should be also noted that the error bar of the data points is large.
Anyway, although the $L_0 - t_b - E_{p,i}$ correlation alone can not be used to constrain the cosmological parameters accurately,
it can be used as a new way and a useful supplement to the current widely used cosmological probes.

\begin{table*}[htbp]
  \centering
  \caption{Constraints on the cosmological parameters for different universe model and GRB samples}
    \begin{tabular}{c|c|c}
  \hline
  \hline
    Flat $\Lambda$CDM model   & $\Omega_m$ \\
  \hline
  Optical sample & $0.697_{-0.278}^{+0.402}$ \\
  X-ray + Optical sample & $0.313_{-0.125}^{+0.179}$ \\
  \hline
  Non-flat $\Lambda$CDM model   & $\Omega_m$ & $\Omega_{\Lambda}$\\
  \hline
  Optical sample & $0.713_{-0.278}^{+0.346}$ & $0.981_{-0.580}^{+0.379}$\\
  X-ray + Optical sample & $0.344_{-0.112}^{+0.176}$ & $0.770_{-0.416}^{+0.366}$ \\
  \hline
  \hline
    \end{tabular}%
  \label{tab:cos_para}%
\end{table*}%

\section{Conclusions}

We have investigated whether the three-parameter correlation in GRBs, $L_0-t_b-E_{\gamma,iso} (E_{p,i})$,
can be used to constrain the cosmological parameters. We found that $L_0-t_b-E_{p,i}$ correlation is better than $L_0-t_b-E_{\gamma,iso}$
for limiting the cosmological parameters. We have selected two groups of GRB samples:
one group is composed of 31 long GRBs with X-ray plateau followed by a normal decay phase with a decay index of about -2.
The other group is comprised of 50 optical samples with light curves from a shallow decay (or a slight rising phase)
to normal decay (or an even steeper decay). The GRBs selected by classification of decay phase are generally assumed to have the same physical origin.
We have used the GP method to calibrate the three-parameter correlation, and then the cosmological parameters are constrained
by using the selected samples. Employing the optical sample, we get the best-fitting result of the parameter
$\Omega_m = 0.697_{-0.278}^{+0.402}(1\sigma)$ for a flat universe model.
For the non-flat model, the best-fitting results are $\Omega_m = 0.713_{-0.278}^{+0.346}$, $\Omega_{\Lambda} = 0.981_{-0.580}^{+0.379}(1\sigma)$. Utilizing all the optical and
X-ray sample, for a flat $\Lambda$CDM model, the best-fitting matter density parameter is
$\Omega_m = 0.313_{-0.125}^{+0.179}(1\sigma)$, and
$\Omega_m = 0.344_{-0.112}^{+0.176}$, $\Omega_{\Lambda} = 0.770_{-0.416}^{+0.366}(1\sigma)$ for a non-flat $\Lambda$CDM model.

Based on the results, we found that the constraints on the cosmological parameters obtained from GRBs of
X-ray and optical samples are weaker than those obtained from SNe Ia. Therefore, $L_0 - t_b - E_{p,i}$ correlation can not simply
be used to accurately measure the cosmological parameters at present, but can be seen as a supplement of cosmological probes.
We also found that the final constraints from the X-ray sample are not as tight as that of the combined sample, and
there is a relatively large dispersion for the $L_0 - t_b - E_{p,i}$ correlation. Our results support that selecting
GRB samples from possible identical physical mechanism is crucial for cosmological purposes.

In the future, it is expected that
more sophisticated multiband detectors can detect much more high-redshift GRBs, which will help us better study GRBs and
the high-redshift universe. In addition, the corresponding parameters, e.g., $L_0$, $t_b$ and $E_{p,i}$, can be better determined,
and then the $L_0 - t_b - E_{p,i}$ correlation could be used to constrain the cosmological parameters better.

\section{Acknowledgments}
We thank the referee for helpful comments. This work is supported by the National Natural Science Foundation of
China (Grant Nos. U2038106, 12273009), Shandong Provincial Natural Science Foundation (ZR2021MA021), Jiangsu Program for Excellent Postdoctoral Talent (20220ZB59) and China Postdoctoral Science Foundation (2022M721561).

\end{document}